\pdfoutput=1 
\documentclass[twoside]{IEEEtran}
\usepackage[noadjust]{cite}

\ifCLASSINFOpdf
   \usepackage[pdftex]{graphicx}
\else
\fi
\usepackage{amsmath}
\usepackage{array}

  \usepackage[caption=false,font=footnotesize]{subfig}
\usepackage{url}
\usepackage{multirow}

\usepackage{balance}

\begin{document}
%
\title{Self-Supervised Learning of Perceptually Optimized Block Motion Estimates for Video Compression}

%
%
%
\author{Somdyuti Paul,
        Andrey Norkin,
        and Alan C. Bovik

\thanks{Somdyuti Paul and Alan C. Bovik are with the Department
of Electrical and Computer Engineering, University of Texas at Austin, Austin,
TX, 78712 USA (email: somdyuti@utexas.edu, bovik@ece.utexas.edu).}
\thanks{Andrey Norkin is with Netflix Inc. Los Gatos, CA, 95032 USA (email: anorkin@netflix.com).}
\thanks{This work is supported by Netflix Inc.}}

%
%

\markboth{IEEE Transactions on Image Processing}%
{Paul \MakeLowercase{\textit{et al.}}: Self-Supervised Learning of Perceptually Optimized Block Motion Estimates for Video Compression}

%

%
\IEEEpubid{\begin{minipage}{\textwidth}\ \\[12pt] \\
\copyright 2022 IEEE. Personal use of this material is permitted.  Permission from IEEE must be obtained for all other uses, in any current or future media, including reprinting/republishing this material for advertising or promotional purposes, creating new collective works, for resale or redistribution to servers or lists, or reuse of any copyrighted component of this work in other works.”\\ 
\end{minipage}} 


\maketitle


\begin{abstract}
Block based motion estimation is integral to inter prediction processes performed in hybrid video codecs. Prevalent block matching based methods that are used to compute block motion vectors (MVs) rely on computationally intensive search procedures. They also suffer from the aperture problem, which tends to worsen as the block size is reduced. Moreover, the block matching criteria used in typical codecs do not account for the resulting levels of perceptual quality of the motion compensated pictures that are created upon decoding. Towards achieving the elusive goal of perceptually optimized motion estimation, we propose a search-free block motion estimation framework using a multi-stage convolutional neural network, which is able to conduct motion estimation on multiple block sizes simultaneously, using a triplet of frames as input. This composite block translation network (CBT-Net) is trained in a self-supervised manner on a large database that we created from publicly available uncompressed video content. We deploy the multi-scale structural similarity (MS-SSIM) loss function to optimize the perceptual quality of the motion compensated predicted frames. Our experimental results highlight the computational efficiency of our proposed model relative to conventional block matching based motion estimation algorithms, for comparable prediction errors. Further, when used to perform inter prediction in AV1, the MV predictions of the perceptually optimized model result in average  Bj{\o}ntegaard-delta rate (BD-rate) improvements of  \mbox{-1.73\%} and -1.31\%  with respect to the MS-SSIM and  Video Multi-Method Assessment Fusion (VMAF) quality metrics, respectively, as compared to the block matching based motion estimation system employed in the SVT-AV1 encoder. 
\end{abstract}


%
\IEEEpeerreviewmaketitle

\section{Introduction}
\label{sec:intro}
%
%
%
%
\IEEEPARstart{B}{lock} based motion estimation and compensation techniques are crucial to systems that perform such diverse tasks as video coding, frame rate up-conversion, 3D scene reconstruction etc. Motion estimation plays a key role in block based hybrid video codecs by facilitating the process of inter prediction, which reduces temporal redundancies inherent in natural videos. In fact, a block based hybrid encoder derives much of its compression capability from the inter prediction process. The use of a wide range of variable block sizes in modern codecs has further increased the complexity of the motion estimation process as compared to legacy codecs that used block sizes from a more limited range. Such codecs thereby avoid the extreme expense of exhaustive search when estimating block motion vectors (MVs), by instead using fast block matching algorithms that greatly reduce computational complexity at the expense of higher prediction errors. However, optimization of the motion estimation process still largely relies on traditional pixelwise error (match) criteria, such as the sum of squared errors (SSE) or the sum of absolute differences (SAD). Simple pointwise measures of signal fidelity like these are unreliable predictors of perceptual picture quality \cite{mse}, which is the visual attribute signal processing systems ultimately should optimize. Thus, it is quite plausible that the  efficiency of the motion estimation modules used in current video encoders could be improved by incorporating measures of perceptual quality into their design. 

\begin{figure*}
\centering
    \includegraphics[width=17cm]{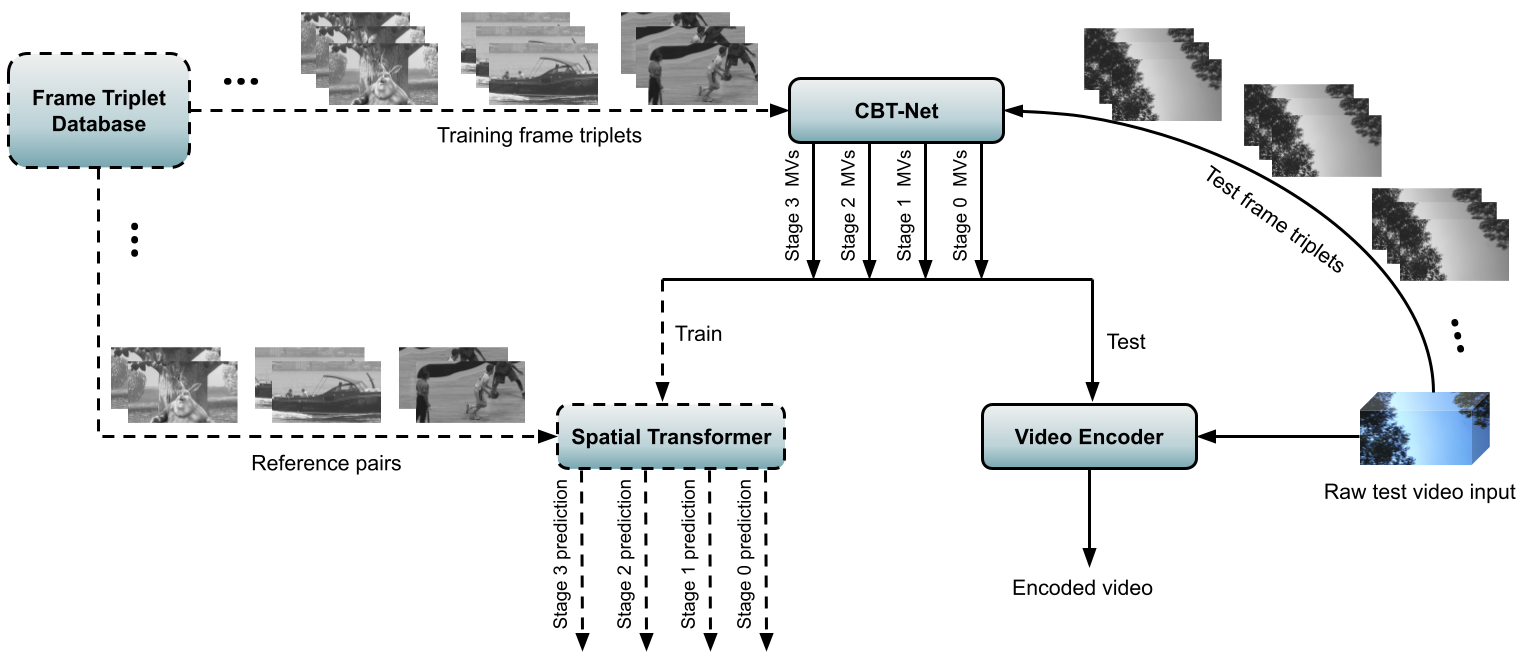}
    \caption{\small{{Flow diagram of the proposed MV estimation framework.}}}
    \label{fig:overview}
\end{figure*}

\IEEEpubidadjcol While significant research efforts have been directed towards improving search based fast block matching algorithms \cite{diamond, arps}, the premise of using data driven deep learning techniques to learn block MVs has not received much research attention, despite successes attained on the related problem of dense optical flow estimation. Here, we address the question of perceptual optimization of block motion estimation, by introducing a data-driven, self-supervised framework for learning block translations for video encoding. We show that our method can be directly used in the inter prediction step of an AV1 encoder \cite{av1, av1journ}, the latest royalty free video coding format. A multi-stage convolutional neural network (CNN) model is used to predict  MVs for all  block sizes required for AV1 inter-prediction. This composite block translation network, which we refer to as CBT-Net, is designed to predict the MVs of bidirectionally predicted frames (B-frames) using a frame triplet as its input. Each input triplet comprises a B-frame that is to be predicted through motion compensation, along with past and future reference frames.  At each of its output stages, the model predicts MVs of blocks of a certain size, starting from the largest blocks at the first stage. Motion compensated frame predictions are generated using a spatial transformer module \cite{stn}, allowing the model to be trained in a self-supervised manner, using the multi-scale structural similarity (MS-SSIM) measure \cite{ms-ssim} which is a widely used, accurate and differentiable measure of perceptual visual quality. In this way, the trained CBT-Net is optimized for MV estimation, and used to replace the search based motion estimation of the SVT-AV1 encoder \cite{svt, svtpaper, svtblog}. We show that making this substitution results in significant perceptual rate-distortion (RD) gains when encoding AV1 bitstreams. The overall proposed MV estimation framework is schematically outlined in Fig. \ref{fig:overview}. 

We begin by noting that it is difficult to apply objective perceptual quality models like MS-SSIM to optimize block motion estimation algorithms used in video compression, since objective measurements of perceptual quality are less meaningful on the relatively small prediction block sizes that remain prevalent in current video codecs, alongside larger prediction block sizes. However, using CNN based model design, we are able to maximize the perceptual quality on entire predicted frames, instead of minimizing pixelwise prediction errors on individual blocks. This makes it possible to exploit differentiable perceptual quality models, like SSIM or MS-SSIM in the task of motion estimation, leading to improved RD performance not only against MS-SSIM, but also with respect to the equally widely-used perceptual Video Multi-Method Assessment Fusion (VMAF) \cite{vmaf} algorithm, and even against the traditional peak signal-to-noise ratio (PSNR). The primary contributions of our work can be summarized as follows:
\begin{enumerate}
\item We developed a composite model that learns to predict MVs over multiple block sizes in a hierarchical fashion, eliminating search processes required in previous block matching based ME approaches. 
\item We exploited the spatial transformer module to achieve block translations in a continuous and learnable manner. 
\item As an example, we substituted the MVs computed using hierarchical motion estimation and integer full search in the SVT-AV1 encoder with the trained CBT-Net predictions, and showed that it leads to a significant improvement in the RD performance.
\end{enumerate}

We selected the SVT-AV1 encoder as the baseline since it is an open source AV1 encoder which performs well in terms of both compression efficiency and computational complexity, and since its motion estimation process uses the original video frames as reference frames rather than decoded ones (open loop motion estimation) for deriving integer pixel precision MVs, and thus fits our proposed motion estimation framework. Conversely, the standardized reference encoders generally deploy closed loop motion estimation architectures, where the decoded and reconstructed frames are used as reference frames. Nevertheless, many emerging practical encoders adopt the open loop architecture due to the advantage it imparts in terms of efficient parallelization and real-time processing performance, while the sequential nature of a typical closed loop architecture limits the parallelization  that can be achieved. In SVT-AV1, only the fractional-pel refinement stage that follows the open loop integer precision MV estimation takes place in closed loop, as explained in Section \ref{subsec:integration}. Further, the open loop motion estimation architecture substantially simplifies the process of generating the reference frames necessary to train data driven motion estimation models, since the source videos need not be encoded and decoded to generate reference frames as in the case of a closed loop architecture.
Thus, our motion estimation framework can also be directly applied to other practical and scalable encoders such as SVT-HEVC \cite{svthevc} and SVT-VP9 \cite{svtvp9}, which use the open loop architecture, while its application to closed loop encoder architectures such as HM \cite{hm} and VTM \cite{vtm} necessitates a different training data generation policy from the one proposed here.

The rest of the paper is organized as follows. We review existing related work in Section \ref{sec:review}. Section \ref{sec: database} describes the frame triplet database which is essential to our self-supervised MV estimation approach. We introduce the proposed framework for learned motion estimation culminating in the CBT-Net model in Section \ref{sec:framework}. Section \ref{sec:results} presents the experimental results, based on which we draw conclusions in Section \ref{sec:conclusion}.

\section{Related Work}
\label{sec:review}

Deep learning based methodologies are being increasingly explored as potentially efficacious alternatives to the traditional block based hybrid video coding framework. End-to-end deep video compression schemes such as \cite{dvc, learned} have been shown to achieve a compression efficiency comparable to or better than mainstream codecs such as HEVC and VP9, adducing to the considerable progress made in this direction. Many end-to-end video compression schemes such as \cite{dvc} have employed CNN models to conduct optical flow estimation that is used to generate temporal predictions, subsequently compressing the motion information and residual signals using learned autoencoders. In \cite{learned} the motion estimation process was generalized to allow for efficient representation of complex motion types that cannot be captured by simple pixel translations, within an end-to-end compression framework. A bidirectional inter frame interpolation technique for video compression was developed in \cite{inter-frame}, where a pretrained optical flow estimation model was combined with a learned encoder-decoder pair to perform temporal prediction, and the residuals were computed in the latent space instead of pixel space to enable the use of existing deep image compression models to jointly encode the latent representations of key frames as well as residuals. Explicit motion estimation was avoided in the video compression framework introduced in \cite{withoutmotion} using displaced frame differences to capture motion regularities, along with a learned encoder-decoder network to achieve spatiotemporal reconstruction. 

Learned video coding tools constitute another salient way in which deep learning based approaches have contributed to the advancement of video compression technology. Several deep learning based coding tools have been developed to optimize specific functions of hybrid video codecs, such as fractional interpolation \cite{fractionalinterpolation}, in loop filtering \cite{inloop,residualinloop}, block partition prediction \cite{hevcpartition, hfcn}, angular intra prediction \cite{angular}, among others, to improve compression efficiency and/or reduce encoding complexity. Nevertheless, data driven learning based solutions to improve block based inter prediction have been less explored, despite this process being a key step in the traditional video coding pipeline. A combination of a fully connected network and a CNN was used to fuse spatial and temporal information for inter prediction to achieve coding gains in \cite{interpred}. Learning based solutions were also developed to reduce  blocking artifacts that often arise from block motion compensated prediction at low bitrates, using a CNN model to non-linearly combine bidirectional predictions in \cite{bi-comp} and to refine predicted blocks in \cite{comp-refine}. 

The veritable improvements that image and video compression systems can derive by exploiting principles of human perception of visual information have been extensively emphasized by researchers \cite{perceptualsurvey}. However, rapid advancement in this direction is often stymied by challenges arising from our limited understanding of the complex phenomena that govern human perception, the computational complexity of models that seek to capture such perceptual effects, and by the intricacies of hybrid codec design that make it difficult to seamlessly incorporate and integrate novel components into the coding pipeline \cite{perceptualcoding}. In \cite{learningjnd}, a deep learning based just-noticeable-quantization-distortion model was developed to eliminate perceptual redundancy in video coding, resulting in substantial coding gains. Properties of the human visual system, such as contrast sensitivity and visual masking, have also been used to enhance the quality of encoded videos by adaptively varying the quantization parameter \cite{aq, cnnaq}. Saliency and visual attention models constitute yet another way of conducting bit allocation to achieve perceptual coding gains \cite{saliencyaware, saliency}. Other methods of perceptual coding have employed reliable objective visual quality models like the structural similarity (SSIM) index \cite{ssim} and VMAF \cite{vmaf}; a SSIM-based divisive normalization scheme was proposed in \cite{perceptualssim} that enhanced the coding efficiency and visual quality of hybrid codecs, while \cite{proxiqa} substituted a proxy for the objective VMAF metric using a CNN regression model to derive a perceptual loss for training an end-to-end deep compression model. 

Several block matching approaches have been developed to improve tradeoffs between the computational complexity of the motion search against the prediction error. A majority of these rely on using efficient search patterns such as diamond search (DS) \cite{diamond} and adaptive rood pattern search (ARPS) \cite{arps}, while few approaches employ learned models. Instead of searching for best matching blocks in pixel space, the authors of \cite{blocknet} proposed block matching in feature space, where convolutional features were extracted from block pixels, and a simple average operator was used to perform representative matching in feature space to reduce the computational complexity of block motion estimation. In \cite{cnn-motion}, a CNN model was trained to predict the similarity of two input blocks, where the trained model was subsequently used to perform block matching for frame interpolation. However, none of these MV estimation methods completely eliminate the search based block matching process. Furthermore, learned approaches like those in \cite{blocknet, cnn-motion} only predict the MVs for a single block size. Consequently, such MV estimation methods are not directly applicable for inter-prediction in hybrid video codecs, which rely on variable block sizes to obtain precise predictions while maintaining bitrate constraints. Additionally, the design of such models does not account for the perceptual quality of the resulting frame predictions. 

The remarkable success of deep CNNs such  as \cite{flownet, spynet, pwc, unsupflow, unflow} on the related problem of dense optical flow estimation motivates our composite model design. Unlike existing models and algorithms that estimate block MVs, our proposed CBT-Net model entirely eliminates the search process required for block matching, and can collectively predict MVs over multiple block sizes required for the inter prediction process. Moreover, our approach also takes cognizance of the overall perceptual quality of the resulting motion compensated frame predictions by optimizing the MS-SSIM values of the predicted frames, rather than relying on perceptually agnostic pixelwise prediction error optimization that has been exclusively used as matching or optimization criteria in previous works.

\section{Construction of Frame Triplet Database}
\label{sec: database}

A substantially large amount of training data is necessary to learn block MVs using a deep neural network. However, generating a large number of exact block MVs corresponding to each prediction block size used by the encoder is an onerous task, due to the extremely high computational complexity of the associated block matching process. Moreover, motion estimation is inherently an ill-posed problem, due to the projection of the three dimensional true motion on a two dimensional plane, and the exact MVs between two frames cannot be theoretically determined without involving additional constraints. This signifies that even exact block matching algorithms can fail to estimate true motion in the presence of effects such as changes in illumination and occlusion. They are also impaired by the \textit{aperture problem} which refers to the ambiguity in inferring the true direction of motion while viewing only a part of a moving object through a small aperture. This problem intensifies at smaller block sizes, as shown in Section \ref{subsec:pred}. Taking into account the shortcomings of block matching based motion estimation, here we instead propose a self-supervised training approach, which does not require exact block MVs as ground truth data for training the CBT-Net model. Instead, triplets of original frames are used to implicitly learn block translations, as elaborated further in Section \ref{sec:framework}. 

\subsection{Source Content}
\label{subsec:source}
In order to create a database of frame triplets to perform self-supervised training of the CBT-Net model, we collected a diverse set of 109 uncompressed video sequences having native resolutions of $1920 \times 1080$ pixels (1080p) or higher from publicly available sources \cite{mcml, sjtu, uvg, netflix-open}. We selected the sources to encompass different motion types, including camera motion, sports content, and computer graphics image (CGI) content, and different framerates ranging from 24 to 120 frames per second. A list of the contents selected for this database is provided in \cite{mv_database}, specifying the sources from which the contents were collected, along with relevant information describing the lengths, original resolutions and framerates of each content. Some of the contents comprised multiple video shots. All sources were converted to 8 bit representation, and the luma channels were extracted. The sources which were at a higher resolution were downsampled to $1920 \times 1080$ using Lanczos resampling, with kernel width $\alpha=3$. If a source content comprised multiple scenes, it was split into two or more segments such that no segment contained any scene changes. 

The sequences used to construct the frame-triplet database include many videos containing camera (ego) motion, as well as combinations of camera and object motions. Several videos were also captured with static cameras (object motion). For example, the training set was constructed using 87 video shots that have both object motion and ego motion, 10 video shots that exclusively contained camera motion over static scenes, and 26 video shots that were captured with static cameras. 

%

\begin{figure}[tb]
\centering
\includegraphics[width=9cm]{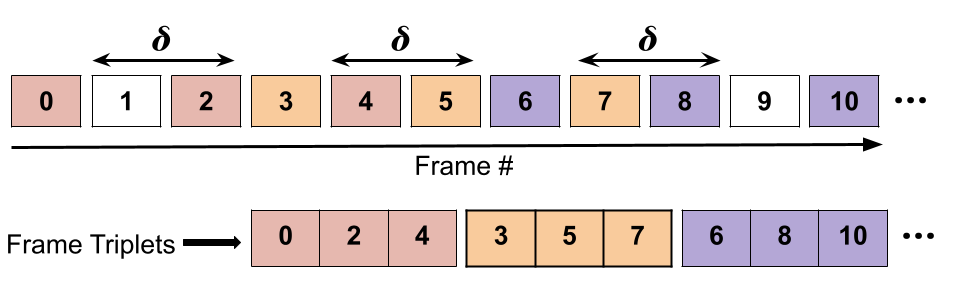}
\caption{\small{{Frame triplet extraction scheme used when creating the database (at layer 3).}}}
\label{fig:triplet}
\end{figure}

\subsection{Frame Triplet Extraction}

\begin{figure}[tb]
\centering
\includegraphics[width=8.8cm]{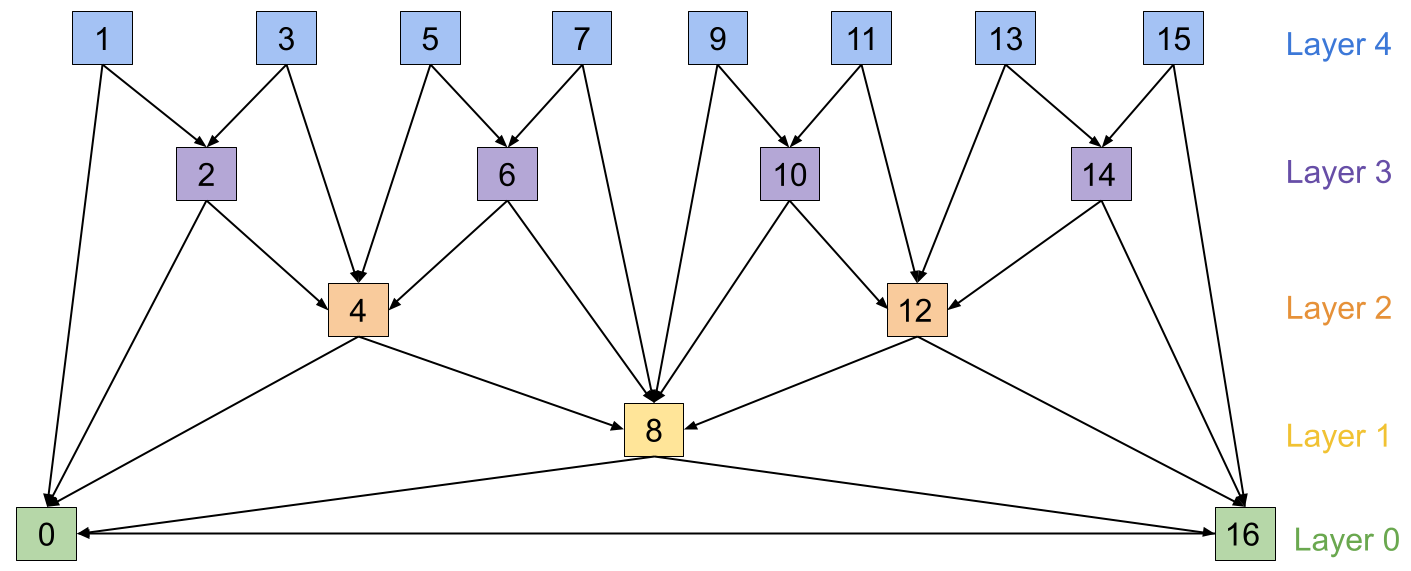}
\caption{\small{{Five-layer hierarchical temporal prediction structure used in SVT-AV1. }}}
\label{fig:temporal_layers}
\end{figure}

Contemporary hybrid codecs typically use hierarchical prediction structures to conduct bidirectional inter prediction. A group of pictures (GOP) is subdivided into mini-GOPs, where the mini-GOP size is the number of frames between two consecutive unidirectionally predicted frames within a GOP. For example, a five-layer hierarchical temporal structure with a mini-GOP size of 16, as used in the SVT-AV1 encoder, is illustrated in Fig. \ref{fig:temporal_layers}, where the top four layers correspond to B-frames; each B-frame uses a pair of reference frames, one aligned along each temporal direction. We used this prediction structure to guide our frame triplet database creation. Specifically, we extracted frame triplets to correspond to B-frames present at the top four temporal layers, such that the distances between the B-frames and their references at each layer are the same as those shown in Fig. \ref{fig:temporal_layers}. 

Let $(R_P, Q, R_F)_k$ constitute a frame triplet at layer $k$ for $k=1 \cdots 4$, where $Q$ denotes the intermediate frame that is to be predicted using block MVs computed with respect to a past reference frame $R_P$ and a future reference frame $R_F$. Let $d(F_i, F_j)$ denote the number of frames between any two frames $F_i$ and $F_j$ of a video sequence, i.e. the distance between the two frames. A set of frame triplets extracted at each temporal layer can then be represented as:
\begin{equation}
\mathcal{S}_k = \{(R_P, Q, R_F)_k: d(R_P,Q)=d(R_F,Q)=2^{4-k} \} \text{,}
\label{eq:database}
\end{equation}
for $k=1 \cdots 4$. Since the prediction structure shown in Fig. \ref{fig:temporal_layers} is symmetric, $d(R_P, Q)=d(R_F, Q)$ $\forall k$. Our database $\mathcal{S}$ is thus composed of four sets of frame triplets described by (\ref{eq:database}), i.e. $\mathcal{S} = \{\mathcal{S}_1, \cdots, \mathcal{S}_4\}$.

We extracted frame triplets from each source content according to (\ref{eq:database}) to constitute each set $\mathcal{S}_1, \cdots, \mathcal{S}_4$. To reduce the temporal redundancies between adjacent triplets extracted from the same content, we imposed a gap of $\delta$ frames between adjacent triplets, as illustrated in Fig. \ref{fig:triplet} using layer 3 as an example. We used $\delta = 2$ frames for layers 1, 2 and 3, and $\delta = 3$ frames for layer 4. The sets $\mathcal{S}_1, \cdots, \mathcal{S}_4$ are individually partitioned into training and validation sets, with distinct contents in each partition, as summarized in Table \ref{table:mvdatabase}. The contents were divided between the training and validation partitions, such that each partition contains a range of different motion intensities and content types, such as CGI and sports contents. The contents included in each partition are also separately listed in \cite{mv_database}.

\begin{table}
  \caption{Summary of frame triplet database for MV estimation.}
  \centering
  \begin{tabular}{|c|c|c|c|c|c|c|c|c|c|c|}
    \hline
    \multirow{2}{1.2cm}{\hfil \textbf{Partition}} &  \multirow{2}{1cm}{\textbf{\# of contents}} & \multicolumn{4}{c|}{\textbf{\# of triplets samples for each layer}} \\
    \cline{3-6}
    & & $\mathcal{S}_1$ &  $\mathcal{S}_2$  & $\mathcal{S}_3$ & $\mathcal{S}_4$  \\
    \hline
     \hfil Training & 97 & 13,943 & 16,959 & 17,169 &  13,166 \\ \hline
     \hfil Validation & 12 & 1308 & 1361 & 1388 & 1058 \\
\hline
  \end{tabular}
  \label{table:mvdatabase}
\end{table}

\begin{figure*}
\centering
\includegraphics[width=18cm]{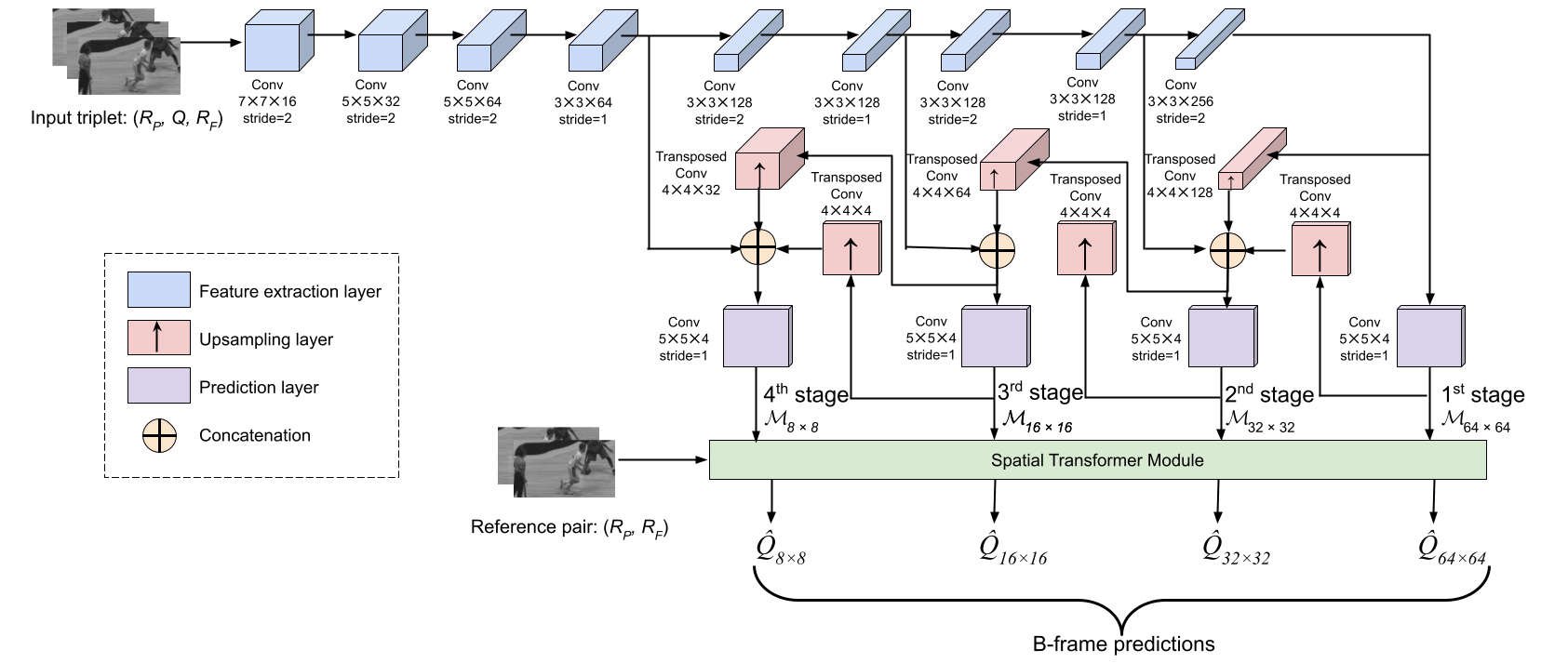}
\caption{\small{{Architecture of the multi-stage CBT-Net model, along with the spatial transformer module used to generate frame predictions.}}}
\label{fig:mv_model}
\end{figure*}

As the distance between a B-frame to be predicted and its reference frames progressively increases from the top towards the bottom of the hierarchical prediction structure, it is reasonable to expect that the amounts of block displacements might differ among the temporal layers, with lower temporal layers being associated with larger block displacements. Thus, the sets $\mathcal{S}_1, \cdots, \mathcal{S}_4$ whose elements conform to the reference-to-predicted frame distances specified by the target prediction structure of Fig. \ref{fig:temporal_layers}, provide a systematic way of adequately representing the distributions of block displacements at each temporal layer. By independently training a separate instance of the CBT-Net model with the frame triplets from each of the four sets, the disparate amounts of block displacements corresponding to the different temporal layers can be effectively captured. Although the frame triplets comprising the database can also be extracted by following other policies, such as randomly choosing the distance between a B-frame and its references (within the limit of the mini-GOP size), to train a single instance of the model for all temporal layers, dividing the database among the four temporal layers, allows for more efficient training, since the separate instances of the model can be trained simultaneously with their corresponding datasets. 

\section{Proposed Method}
\label{sec:framework}

While our method is general, and could be applied to any number of prediction layers and different possible block sizes, we continue to exemplify our method by applying it in the SVT-AV1 framework. The open loop motion estimation process of the SVT-AV1 encoder uses four different prediction block sizes ($64 \times 64$, $32 \times 32$, $16 \times 16$ and $8 \times 8$) to generate inter predictions. Accordingly, we designed the CBT-Net model to concurrently predict the MVs of all non-overlapping $64 \times 64$, $32 \times 32$, $16 \times 16$ and $8 \times 8$ blocks of the input B-frame with respect to both past and future reference frames using a multi-stage process. The frames comprising the input triplet are padded with zeroes such that each frame contains an integral number of blocks of each size. Thus, each input triplet $(R_P, Q, R_F)$ is of dimension $W \times H \times 3$, where $W$ and $H$ are the padded spatial width and height, respectively, that are divisible by 64. 

Traditional MV estimation algorithms perform block matching between a pair of frames. However, using a triplet of frames as input to the CBT-Net model endows it with a notable computational advantage over using frame pairs for motion estimation. Using the frame pairs $(R_P, Q)$ and $(R_F, Q)$ as two separate inputs to the model to compute the forward and backward MVs, would require that all of the features of a B-frame $Q$  be computed for each pair. Using the triplet $(R_P, Q, R_F)$ as the model input instead expedites processing, since the features are extracted from $Q$ only once, and are used to compute both the forward and backward MVs in the same forward pass. Indeed, we have observed $\sim$30\% reduction in inference time using this strategy, with a marginal decline in prediction performance, as compared to using frame pairs as the model input.

\subsection{CBT-Net Architecture}
\label{subsec:arch}

The architecture of the proposed CBT-Net is shown in Fig. \ref{fig:mv_model}, including the spatial transformer module that is used to translate blocks of the reference frames, using the predicted MVs to generate B-frame predictions. As Fig. \ref{fig:mv_model} depicts, this model has four prediction stages, where the MVs of non-overlapping square blocks, of one of the four different sizes is predicted at each stage. The MVs of the largest blocks of size $64 \times 64$ that are used for inter prediction by the SVT-AV1 encoder are predicted at the first stage of the CBT-Net model, followed by the MV predictions on progressively smaller blocks at the succeeding stages. Our motivation for this multi-stage prediction arises from the fact that larger blocks typically contain more distinctive features, making it easier for a model to learn good matches against  blocks of the same size present in other frames. Conversely, on smaller blocks which contain fewer features, the \textit{aperture problem} is exacerbated, rendering the block matching problem more ambiguous. With this in mind, the MVs of larger blocks can be used to guide the MV estimation of co-located smaller blocks in a coarse-to-fine manner, thereby helping to alleviate the aperture problem. Thus, each stage of the motion estimation process, from the second stage onward, relies on both the MVs estimated at the preceding stages, as well as on the feature maps from the preceding stages to make its predictions. 

The CBT-Net model has three types of layers with distinct functionalities, as indicated in Fig. \ref{fig:mv_model}. The \textit{feature extraction layers} are the convolutional layers that extract convolutional features from the input frame triplet. There are nine feature extraction layers in the CBT-Net model, each followed by the customary batch normalization and non-linearity operation, where the latter is implemented using rectified linear units (ReLU). A $5 \times 5 \times 4$ convolutional layer is used as the last layer of each stage to act as the \textit{prediction layer} that produces the MV predictions for that stage using the feature map it receives as input. By design, the first two channels of the output of each prediction layer correspond to the MVs of the past reference $R_P$, while the last two channels correspond to the MVs of the future reference $R_F$. The prediction layer of the first stage uses the feature map generated by the last feature extraction layer of the model, to predict translations of the $64 \times 64$ blocks of the two reference frames $R_P$ and $R_F$ with respect to the frame $Q$. In the second stage, the feature map used as the input to the prediction layer of the first stage, along with the MV predictions from the first stage are individually passed to \textit{upsampling layers} that upsample them by a factor of two, using transposed convolutions \cite{flownet}. These two upsampled volumes and the convolutional feature map of the eighth feature extraction layer are concatenated and used as the input to the prediction layer of the second stage, to generate the MV predictions on $32 \times 32$ blocks. The MV outputs of the remaining stages are predicted by similarly concatenating the feature map of that stage with the upsampled versions of the feature map and the MV predictions from the previous stage, followed by applying the concatenated volume as input to a prediction layer, as shown in Fig. \ref{fig:mv_model}. Thus, by effectively re-using the motion information from the previous stages, the CBT-Net is able to predict the MVs of multiple block sizes, while incurring very little additional computational overhead as compared to traditional block motion estimation algorithms except the hierarchical ones, where a new search based block matching process must be performed for each different block size. 

The computational complexity of conventional block matching based motion estimation algorithms increases as the search range used for matching is increased. However, limiting the search range to small values for the sake of computational tractability precludes accurate motion estimation, especially in the presence of fast moving objects in the scene. By contrast, the effective search range of our proposed approach is limited by the receptive field size of the last feature extraction layer of the model. Thus, the effective search range of the model can be increased by using more feature extraction layers, or by increasing the stride of the convolutional filters at each layer, allowing larger motions to be effectively captured with  lower additional computational overhead, as compared to traditional methods. The size of the effective receptive field of the last layer of the CBT-Net model depicted in Fig. \ref{fig:mv_model} is 255, as shown in Appendix A. Consequently, the effective search range of the model is $\pm 127$ (even though no traditional search is conducted), and the outputs of the prediction layers at each stage were also clipped to lie in this range to give the composite output of the CBT-Net model, denoted by a set of four matrices $\{\mathcal{M}_{64 \times 64}, \mathcal{M}_{32 \times 32}, \mathcal{M}_{16 \times 16}, \mathcal{M}_{8 \times 8}\}$. If the spatial resolution of the input triplet is $W \times H$, then $\mathcal{M}_{S \times S}$, the output MV matrix for $S \times S$ blocks is of dimension $W/S \times H/S \times 4$. 

\begin{figure}[htb]
\centering
\includegraphics[width=9.0cm]{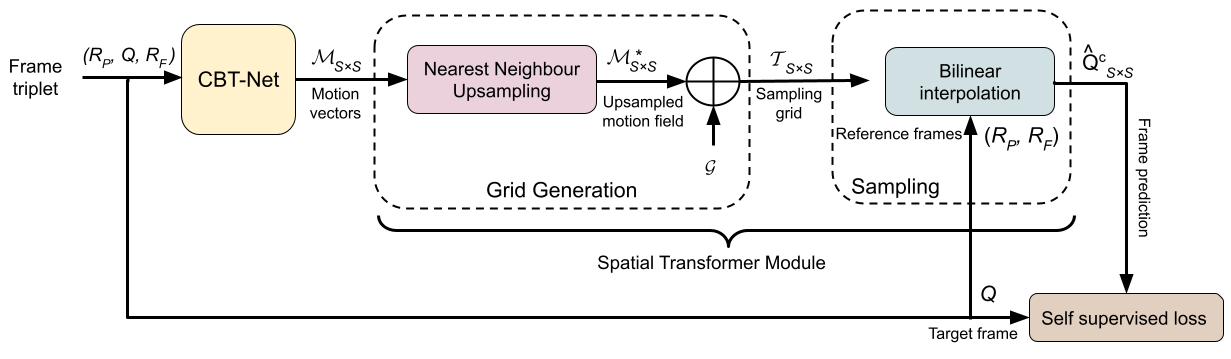}
\caption{\small{{Components of the spatial transformer module.}}}
\label{fig:stn}
\end{figure}

\begin{figure*}[htb]
\centering
\includegraphics[width=18cm]{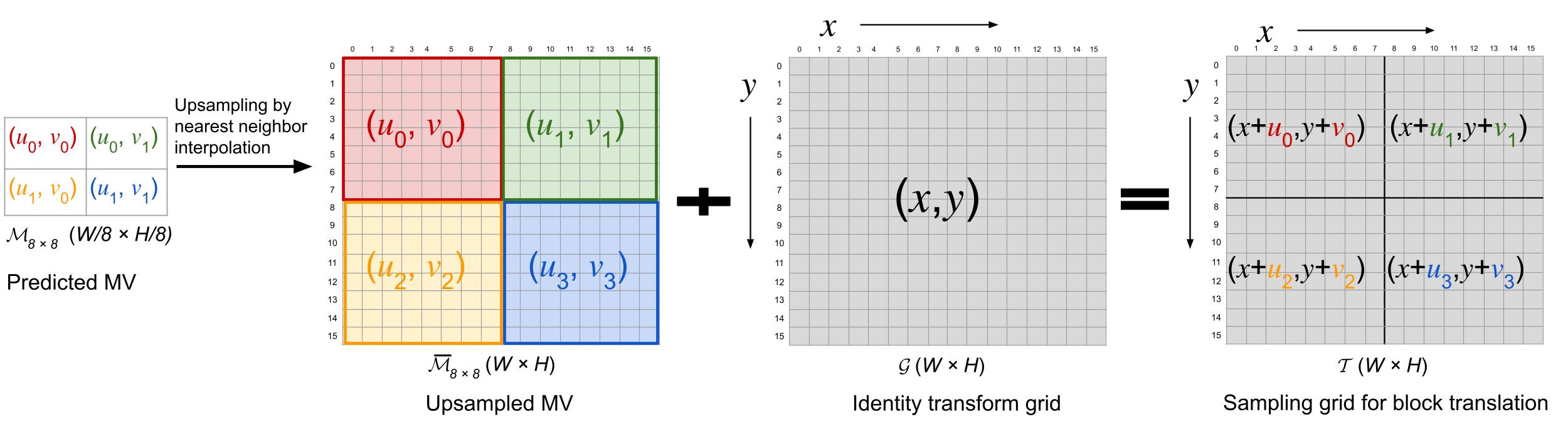}
\caption{\small{{Illustration of the grid generation process for blocks of size $8 \times 8$ and $W=H=16$, where a small spatial dimension is chosen for ease of illustration, and $u_k$ and $v_k$ are the vertical and horizontal components of the estimated MV of the $k^{\text{th}}$ block with respect to a single reference frame.}}}
\label{fig:grid}
\end{figure*}

\subsection{Spatial Transformer Module}
The CBT-Net model gives real-valued MV predictions denoted by $\mathcal{M}_{S \times S}$. However, the block translations to generate the frame predictions as performed in conventional block motion estimation algorithms would require integer valued MVs since the pixels of the reference frames are only defined at integer locations. Simply rounding the real valued MVs to the nearest integer for the purpose of block translations does not solve the problem since the gradients of the round function are zero almost everywhere which is not amenable to learning using back propagation. We propose a solution that overcomes this difficulty using a spatial transformer module. We designed the spatial transformer module to obtain the prediction signals through block translations in a manner that is conducive to back-propagation. A spatial transformer is a differentiable module that can be inserted within a CNN model to perform generic geometric transformations of an image or feature map \cite{stn}. In the context of motion estimation, spatial transformers have been employed to perform feature warping \cite{pwc} and frame warping \cite{unsupflow, unflow} to learn dense optical flows. It was also used in \cite{affine} to learn affine transformations to conduct whole-frame predictions. Unlike these previous methods, in our model, the spatial transformer module estimates differentiable local block translations, by an efficient grid generation process as described later in this section. Specifically, it serves to translate non-overlapping blocks of the reference frames $R_P$ and $R_F$ according to the predicted set of MVs $\{\mathcal{M}_{64 \times 64}, \mathcal{M}_{32 \times 32}, \mathcal{M}_{16 \times 16}, \mathcal{M}_{8 \times 8}\}$ in a differentiable manner. This module thus generates the predictions of frame $Q$ corresponding to each block size and reference frame in order to compute the model's self-supervised loss. The components of the spatial transformer module as well as its role in the proposed framework are illustrated in Fig. \ref{fig:stn}. The inputs to the spatial transformer module are the set of MVs generated by the CBT-Net and the two reference frames. The two steps that constitute the spatial transformer module are described next: 
\begin{itemize}
\item \textit{Grid generation}: the purpose of this step is to find the locations of an input image or feature map that are to be sampled to generate the spatially transformed output image or feature map \cite{stn}. While the predicted elementwise motion parameters can be directly applied to perform warping through coordinate transformation in \cite{pwc, unsupflow, unflow,affine}, the CBT-Net generates the motion parameters at the block level. The grid generation process employed to accomplish block translations can be elucidated with the help of Fig. \ref{fig:grid} for a single block size and reference frame. The predicted motion fields as shown in  Fig. \ref{fig:grid}(a) are first upsampled to the spatial resolution of the input frames using nearest neighbor interpolation. Fig. \ref{fig:grid}(b) represents the upsampled motion field, where all pixels within the individual $S \times S$ blocks get the same value of the pixelwise MVs as indicated by the different color coded regions; the block colors correspond to the ones used to represent the corresponding MVs in Fig. \ref{fig:grid}(a). This step effectively ensures that all pixels within the same block get the same MV $(u_i, v_i)$, thereby achieving the desired block translations. Let the upsampled MV matrix for $S \times S$ blocks of size  $W \times H \times 4$ be ${\mathcal{M}}_{S \times S}^{\ast}$. If $\mathcal{G}(x,y) = (x,y)$ is the sampling grid for the identity transformation as shown in Fig. \ref{fig:grid}(c), then the sampling grid for translating the $S \times S$ blocks is derived as
\begin{equation}
\begin{split}
\mathcal{T}_{S \times S}(x,y,c) & = \mathcal{G}(x,y) +  {\mathcal{M}}_{S \times S}^{\ast}(x,y, 2c:2c+1) \\
& = \Big( x + \mathcal{M}_{S \times S}^{\ast}(x,y, 2c),   \\
& y + \mathcal{M}_{S \times S}^{\ast}(x,y, 2c+1)\Big)
\end{split}
\end{equation}
where $c=0$ for reference frame $R_P$ and $c=1$ for reference frame $R_F$. Fig. \ref{fig:grid}(d) represents this sampling grid, and when the pixels of the reference frames are sampled according to this sampling grid $\mathcal{T}_{S \times S}$, all pixels within the non-overlapping $S \times S$ blocks are translated  by the amount determined by the block MVs $(u_i, v_i)$. 

\item \textit{Sampling}: since the MV matrix ${\mathcal{M}}_{S \times S}^{\ast}$ is real valued, the locations given by the corresponding sampling grid $\mathcal{T}_{S \times S}$ are as well. We use bilinear interpolation to sample the pixels of the reference frame according to the locations given by  $\mathcal{T}_{S \times S}$ to obtain the corresponding prediction $\hat{Q}_{S \times S}^c$ of frame $Q$ obtained by translating $S \times S$ blocks of the reference frame indexed by $c$. 
\end{itemize}

Thus, for each reference frame, the output of the spatial transformer module is the set of predicted frames $\mathcal{Q}^c=\{\hat{Q}_{64 \times 64}^c, \hat{Q}_{32 \times 32}^c, \hat{Q}_{16 \times 16}^{c}, \hat{Q}_{8 \times 8}^c\}$, which are used to compute the self-supervised loss of the CBT-Net model, as described next. 

\subsection{Loss Function}
Self-supervised learning refers to machine learning schemes where the inputs signals of the models are reused to derive pseudo-labels that act as supervision signals for training the model. Such learning schemes have been proven to be useful to solve such problems where ground truth data is difficult to obtain. This is also true for our present problem as discussed earlier in Section \ref{sec: database}. Thus, the supervision signal for our model is derived using the input signals and the model predictions, resulting in a self-supervised learning scheme, whereby the model's loss is computed between the input frame $Q$, which serves as a pseudo label, and its predictions $\hat{Q}_{S \times S}^c$ corresponding to each reference frame and block size. The predictions $\hat{Q}_{S \times S}^c$  are generated by the spatial transformer module by using the model's MV predictions $\mathcal{M}_{S \times S}$ to translate the blocks of the other two frames of the input triplet, $R_P$ and $R_F$, which serve as the reference frames in either direction. We employed MS-SSIM \cite{ms-ssim} to derive this self-supervised loss when training the CBT-Net model as follows:
\begin{equation}
\mathcal{L} = \sum_{c=0}^{1} \sum_{\hat{Q}_{S \times S}^c \in \mathcal{Q}^c} 10 \times \text{log}\big(1-\text{MS-SSIM}(Q,\hat{Q}_{S \times S}^c)\big)\text{,}
\label{eq:loss}
\end{equation}
where MS-SSIM($Q, \hat{Q}_{S \times S}^c$) is the MS-SSIM score of the prediction $\hat{Q}_{S \times S}^c$ with respect to the original frame $Q$. Thus, by minimizing the negative value of the MS-SSIM score (calculated in dB) between $Q$ and its predictions obtained at each stage of the model as given by (\ref{eq:loss}), the average objective visual quality of the inter frame predictions corresponding to all the block sizes was collectively improved by the training procedure.

\begin{table*}
  \caption{Average prediction errors in terms of MAD values on the validation set using HME \cite{svt}, ARPS \cite{arps} and CBT-Net.}
  \centering
  \begin{tabular}{|c|c|c|c|c|c|c|c|c|c|c|c|c|}
    \hline
    \multirow{2}{2cm}{{Temporal layer \#}} &  \multicolumn{3}{c}{\hfil \textbf{$64 \times 64$} blocks} & \multicolumn{3}{|c}{$32 \times 32$ blocks} & \multicolumn{3}{|c|}{$16 \times 16$ blocks} & \multicolumn{3}{c|}{$8 \times 8$ blocks} \\
     \cline{2-13}
    & HME & ARPS & CBT-Net & HME & ARPS & CBT-Net  & HME & ARPS &  CBT-Net & HME & ARPS & CBT-Net \\
    \hline
     \hfil 1 & 15.60 &  \textbf{10.48} & 11.99 & 13.44 &  \textbf{7.07} & 10.88 & 13.00 & \textbf{5.20} & 7.09 & 11.67 & \textbf{4.08} & 6.17 \\ \hline
     \hfil 2 & 13.68 & \textbf{9.29} & 10.54 & 11.50 & \textbf{6.02} & 7.42 & 11.18 & \textbf{4.25} & 5.80 & 9.95 & \textbf{3.23} & 4.97 \\ \hline
     \hfil 3 & 12.16 & \textbf{8.27} & 9.26 & 10.00 & \textbf{5.13} & 6.20 & 9.78 &  \textbf{3.47} & 4.63 & 8.69 & \textbf{2.55} & 3.85 \\ \hline
     \hfil 4 & 11.60 & \textbf{7.49} & 8.15 & 9.42 & \textbf{4.47} & 5.17 & 9.22 & \textbf{2.91} & 3.67 & 7.93 & \textbf{2.07} & 2.92 \\ \hline

  \end{tabular}
  \label{table:prediction}
\end{table*}

\subsection{Integration with SVT-AV1 Encoder}
\label{subsec:integration}
Block motion estimation takes place in two stages in the SVT-AV1 encoder: the full-pixel open loop ME performed before the main frame encoding process and a sub-pixel closed loop refinement performed alongside the mode decisions process. The open loop motion estimation process as implemented in the SVT-AV1 encoder itself consists of the following steps \cite{svt}:
\begin{enumerate}
\item \textit{Hierarchical Motion Estimation (HME)}: first, a three stage HME is conducted at three different resolutions, corresponding to one-sixteenth resolution, one quarter resolution and the base resolution of the input frames, respectively. The goal of HME is to find the best candidate MV for each $64 \times 64$ block. At each stage of HME, the frame is divided into multiple non-overlapping search areas, which are searched independently to produce multiple MV candidates for each $64 \times 64$ block, using the locations given by the best candidate MVs from the previous stage as the starting points for the searches. The HME process then selects the best MV candidate for each $64 \times 64$ block using the SAD metric as the block matching criterion. 
\item{\textit{Integer full search}}: the best MV candidates for the $64 \times 64$ blocks given by HME are used in this step to find the MVs of all blocks of sizes $64 \times 64, \cdots, 8 \times 8$ that are encompassed by each $64 \times 64$ block. The integer full search uses the displaced $64 \times 64$ block locations given by the MVs computed at the previous HME step as the search center to estimate the MVs of all square blocks within each  $64 \times 64$ block using the SAD metric. This comprises the last stage of the open loop ME process, which refines the HME results to give integer pel MVs. 
\end{enumerate}
We replaced the above two sub-processes of the SVT-AV1 encoder's motion estimation process with the trained CBT-Net model. The MVs estimated by the integer full search process of the encoder are multiplied by a factor of four, in order to use the MVs in the quarter pel refinement process. Thus, in order to make the MV predictions of the CBT-Net suitable for subsequent encoding processes, we scaled them in the same manner, and rounded the scaled MVs to the nearest integer values. The horizontal and vertical components of each MV were then combined to form a single 32 bit unsigned integer in accordance with the integer MV syntax used in SVT-AV1. Each block of size $64 \times 64$ contains 85 blocks  combining all four square prediction block sizes used in SVT-AV1, and the integer motion search process of the encoder collectively estimates the MVs of all 85 square blocks within each $64 \times 64$ block as a single unit. So, we organized the four MV matrices corresponding to the four block sizes into a single matrix of size $2 \times W/64 \times H/64 \times 85$, where the first dimension corresponds to the number of reference frames. Finally, the HME and integer full search processes were disabled, and the combined MV matrices were used to provide the MV information for inter prediction while encoding.

\begin{figure*} [htb]
    \centering
  \subfloat[Frames and color coded MVs for \textit{Tennis} sequence. \label{fig:vis_a}]{%
       \includegraphics[width=0.48\linewidth]{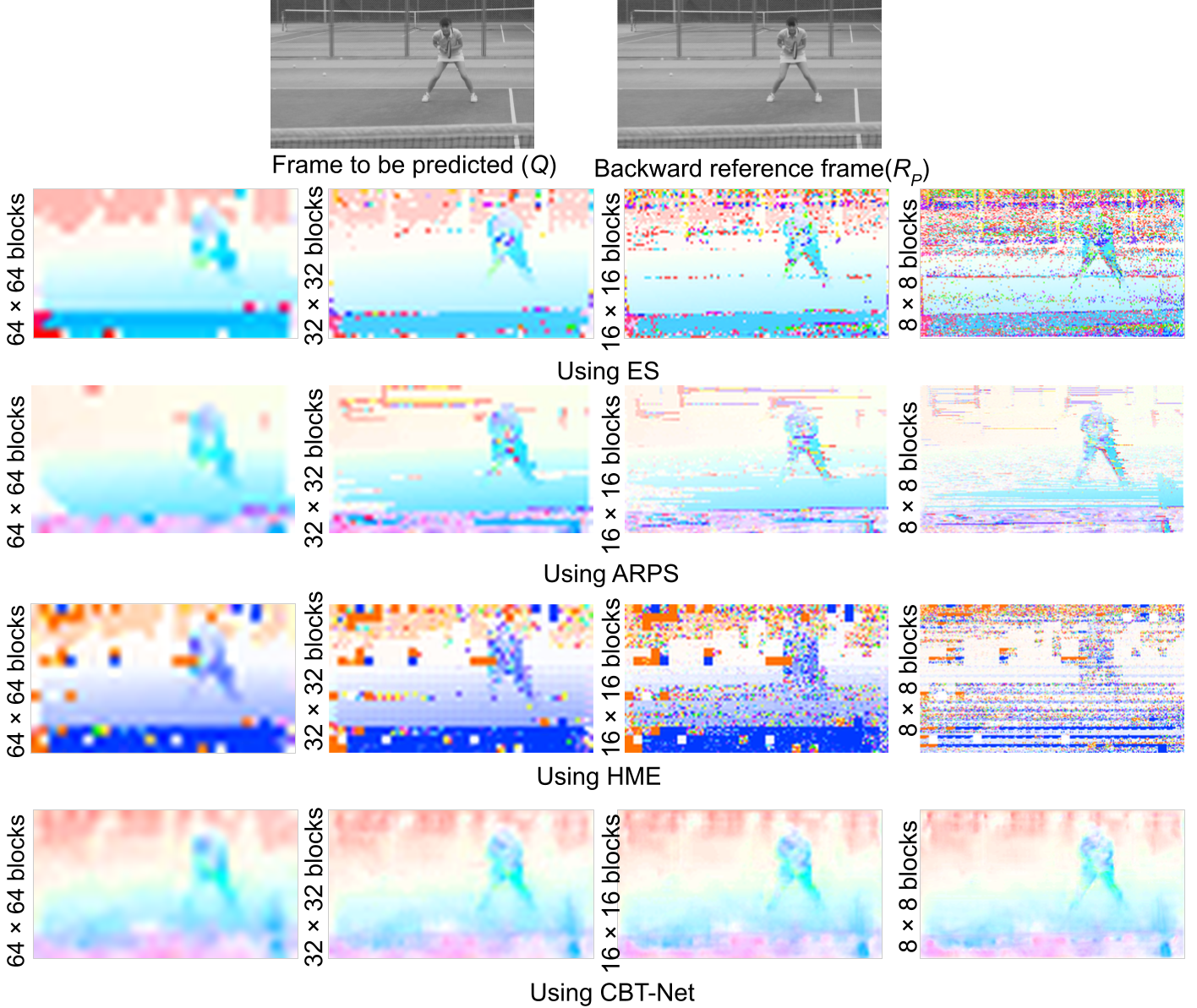}}
       \hspace{0.1cm}
  \subfloat[Frames and color coded MVs for \textit{Netflix Narrator} sequence. \label{fig:vis_b}]{%
        \includegraphics[width=0.48\linewidth]{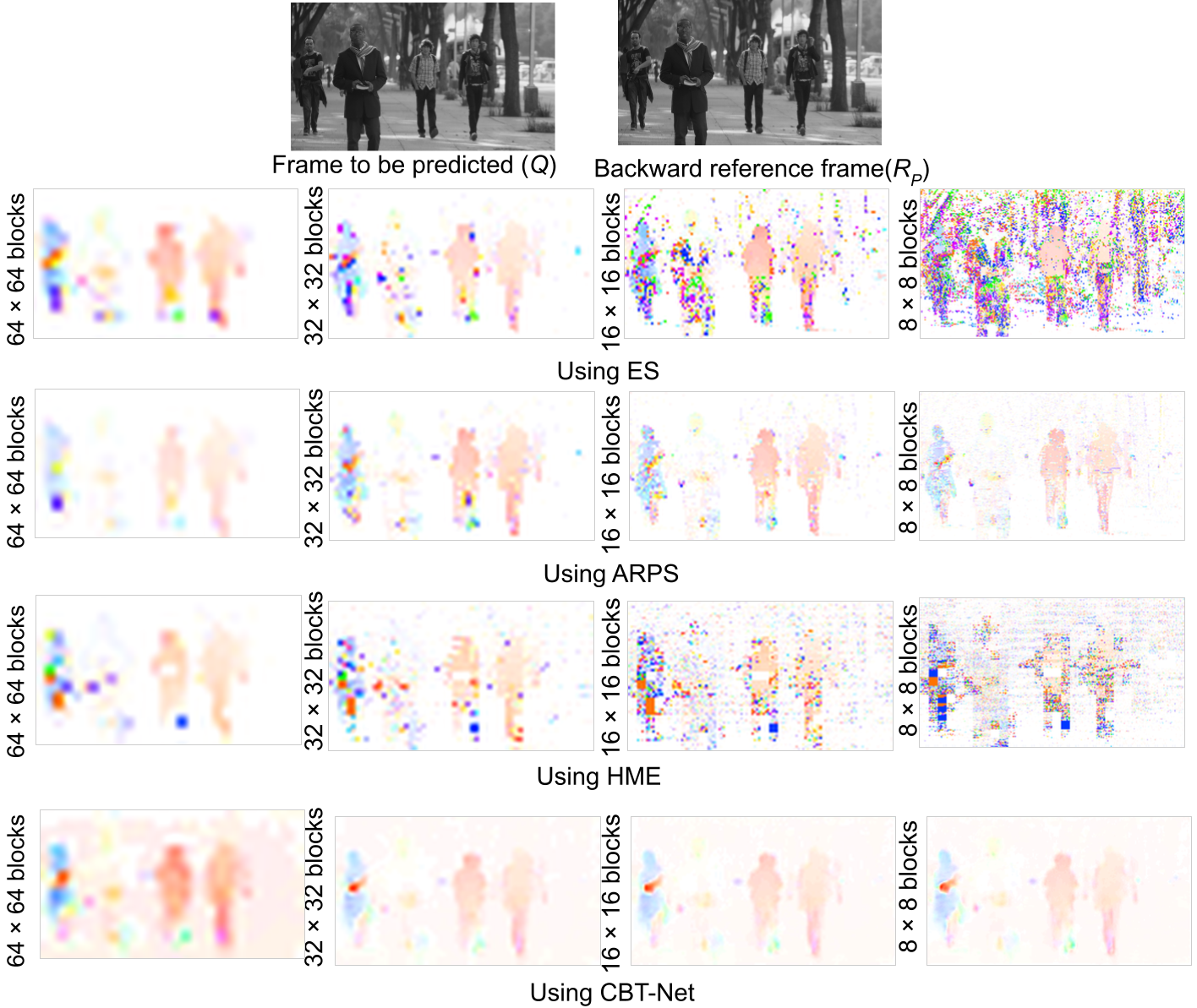}}
  \caption{Visualization of the MVs computed between two frames with the ES, ARPS, HME and CBT-Net model from top to bottom; for each sequence and each method, the MVs computed using block sizes of $64 \times 64$, $32 \times 32$, $16 \times 16$, and $8 \times 8$ shown from left to right.}
  \label{fig:visualization} 
\end{figure*}

\section{Experimental Results}
\label{sec:results}
We trained the CBT-Net model to estimate MVs at two resolutions, $1920 \times 1080$ and $1280 \times 720$. The training data for the smaller of the two resolutions were obtained by downsampling the entire database of frame triplets originally created at $1920 \times 1080$ to $1280 \times 720$ using Lanczos resampling. At each resolution, four different instances of the CBT-Net model were trained for each of the four temporal layers shown in Fig. \ref{fig:temporal_layers}, using the frame triplets from the training partitions of the corresponding sets $\mathcal{S}_1, \cdots, \mathcal{S}_4$. The CBT-Net model was comprehensively evaluated in terms of its prediction error, AV1 encoding performance  achieved using the predicted MVs for inter-prediction, and computational complexity, as explained in Sections \ref{subsec:pred} \ref{subsec:rd}, and \ref{subsec:complexity}, respectively. Although the MV prediction was carried out by the CBT-Net model using luma (Y) channels only, the predicted MVs were used to perform motion compensation of all three channels (Y, U, and V) of the test videos, which were encoded in YUV420p format. 

\subsection{System Settings and Hyperparameters}
The CBT-Net model as shown in Fig. \ref{fig:mv_model} has 1,914,832 trainable parameters. Each instance of the model was trained on four NVidia 1080-TI GPUs, while the prediction and encoding performance evaluation was conducted on a 64 bit, eight core Intel(R) Core(TM) i7-6700K CPU @ 4.00GHz system, running Ubuntu 18.04 (without GPU). All the evaluations conducted utilized all cores of the system. We used the Adam optimizer \cite{adam} with a learning rate of $10^{-4}$ to optimize the parameters of each model instance. The batch size used was 8 in the case of 1080p triplets, and 16 in the case of 720p triplets, respectively.

\begin{figure*}[ht]
    \centering
  \subfloat[1080p sequences\label{1a}]{%
       \includegraphics[width=0.49\linewidth]{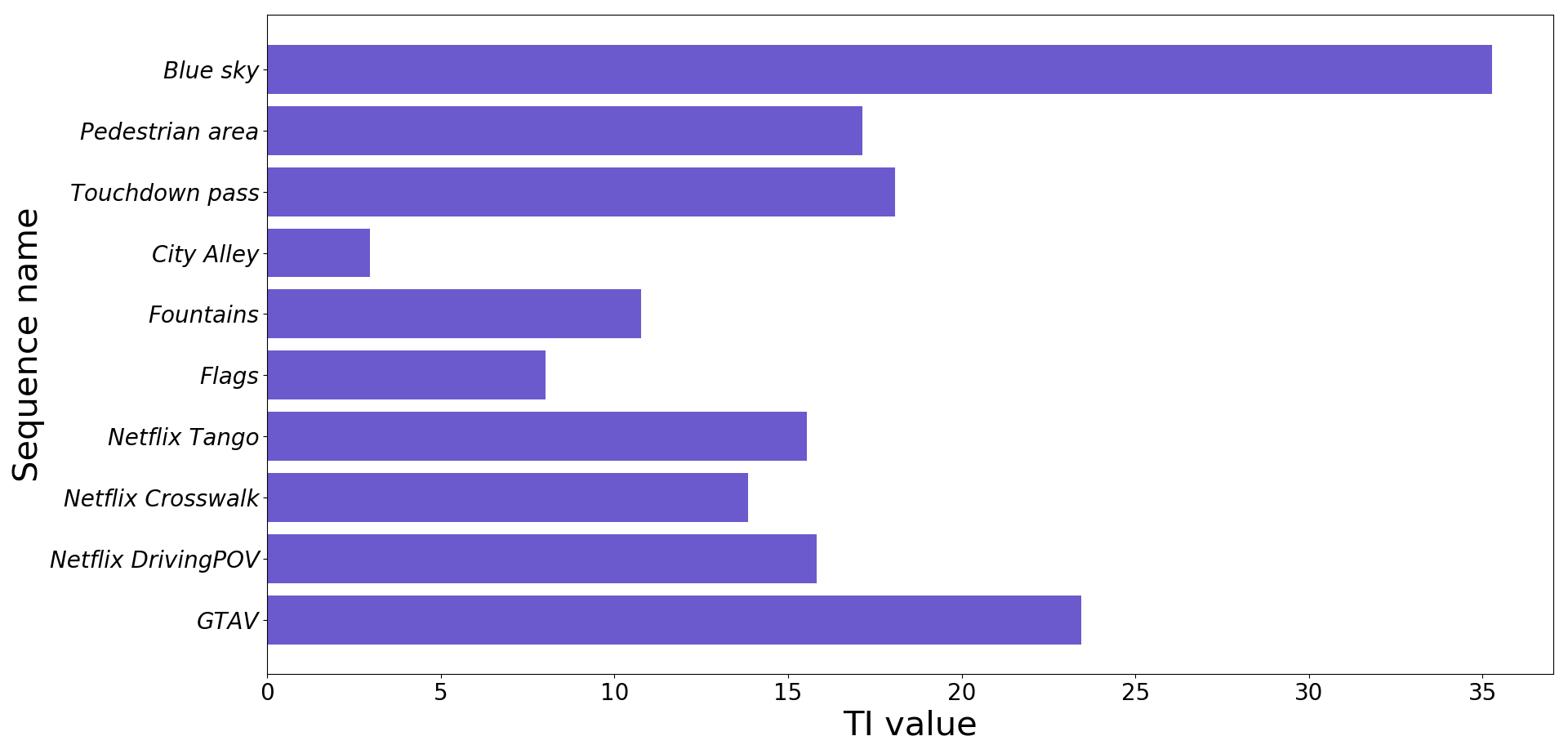}}
    \hfill
  \subfloat[720p sequences \label{1c}]{%
        \includegraphics[width=0.49\linewidth]{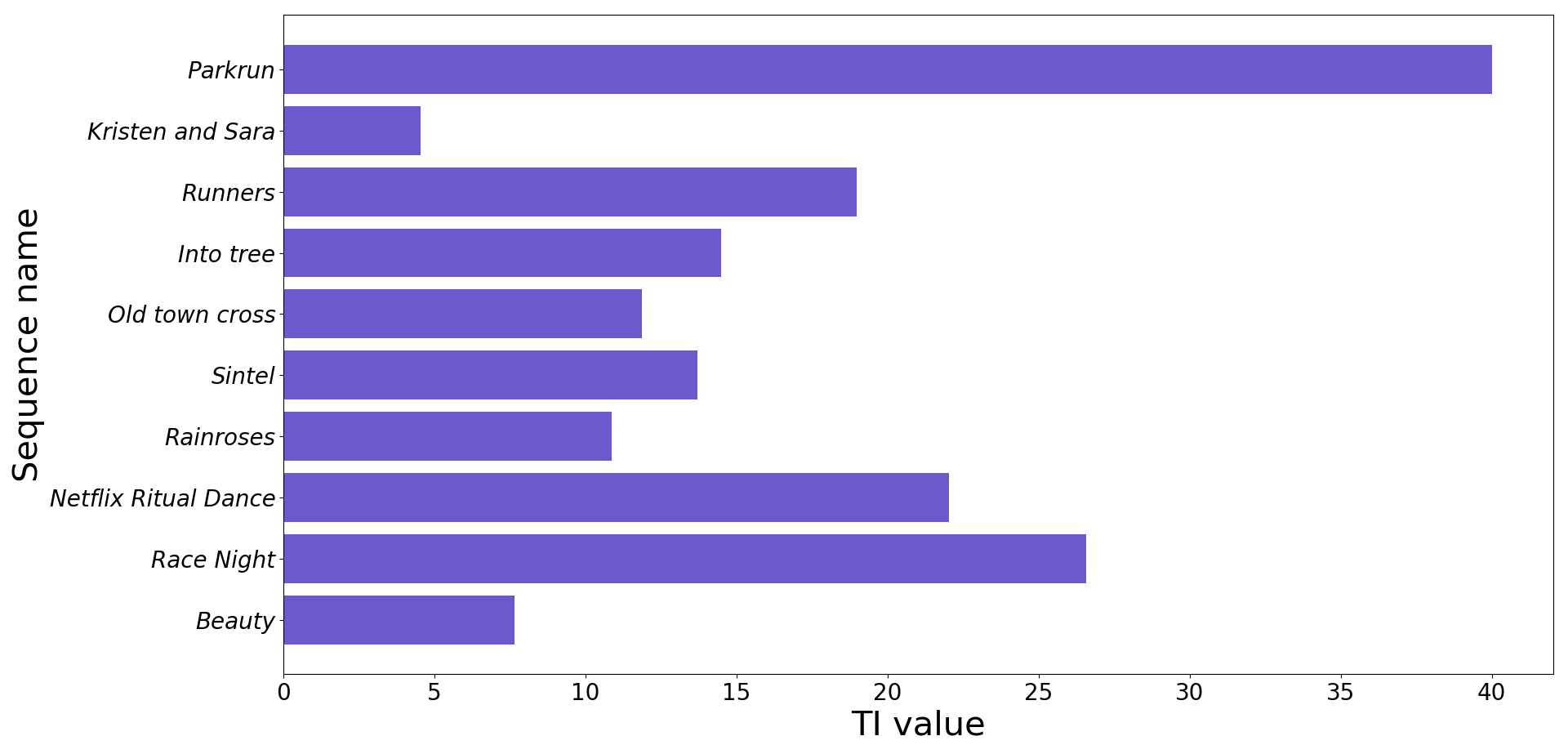}}
  \caption{Distribution of TI values on test sequences at two resolutions.}
  \label{fig:ti_dist} 
\end{figure*}

\begin{table*}[htb]
  \caption{Performance of CBT-Net model and ARPS \cite{arps} with respect to SVT-AV1 baseline.}
  \centering
  \scalebox{0.95}{
  \begin{tabular}{|p{2.5cm}|c|c|cc|cc|cc|cc|}
    \hline
    \multirow{2}{2.7cm}{\textbf{\hfil{Sequence}}} & \multirow{2}{1.5cm}{\textbf{\hfil{Resolution}}} & \multirow{2}{1.5cm}{\textbf{\hfil{\# of frames}}} & \multicolumn{2}{c|}{MS-SSIM BD-rates (\%)}  & \multicolumn{2}{c|}{VMAF BD-rates(\%)} & \multicolumn{2}{c|}{PSNR BD-rates (\%)} & \multicolumn{2}{c|}{$\Delta T$ (\%)}\\
    \cline{4-11}
    &    &   & CBT-Net & ARPS & CBT-Net & ARPS & CBT-Net & ARPS  & CBT-Net & ARPS \\
    \hline
    \hfil{\textit{Blue sky}}  & \hfil{\multirow{3}[20]{*}{1080p}} & 209  & \textbf{-0.95} &  0.21 & \textbf{-2.00} & -0.13  & \textbf{-1.20} &  -0.2 & \textbf{10.2} & 8.4   \\ 
    \hfil{\textit{Pedestrian area}}  &   & 369  & \textbf{-3.22}  & -2.10   &  \textbf{-3.92}  & -2.94 &\textbf{ -2.62}  & -1.48 & \textbf{10.4} & 8.3  \\ 
    \hfil{\textit{Touchdown pass}}  &  & 561  & -6.62  &  \textbf{-6.84} &  -5.10 & \textbf{-5.37} & \textbf{-5.19} & -4.85 & \textbf{10.2} & 8.3 \\ 
    \hfil{\textit{City Alley}}  &   & 593  & -3.39  & \textbf{-5.08}  & -2.42  & \textbf{-3.49} & -1.73 & \textbf{-2.43} & \textbf{5.0} & 1.2 \\ 
    \hfil{\textit{Fountains}}  &  & 289  & 0.30 &  \textbf{-0.04} & \textbf{0.12}  & 0.13 & 0.24 & \textbf{0.13} & \textbf{9.2} & 8.4 \\ 
    \hfil{\textit{Flags}}  &  & 321 &   \textbf{0.41} &  0.44 & 0.49 & 0.49  & \textbf{0.03} & 0.19 & \textbf{9.7} & 7.7 \\ 
    \hfil{\textit{Netflix Tango}}  &  & 289 &\textbf{ -1.48}  & -1.00  & \textbf{-1.03}  & -0.81 & \textbf{-1.22}  & -0.79 & \textbf{11.2} & 9.6 \\ 
    \hfil{\textit{Netflix Crosswalk}}  &  & 289 &  \textbf{-3.79} & -1.77  & \textbf{-3.13}  & -1.49 & \textbf{-2.72}  & -1.51 & \textbf{10.9} & 8.8   \\ 
    \hfil{\textit{Netflix DrivingPOV}}  &  & 289 & -1.84  & \textbf{-1.93}  & \textbf{0.30}  & 1.95 & \textbf{-0.65}  & 0.01 & \textbf{10.8} & 9.9 \\ 
    \hfil{\textit{GTAV}}  &  & 689 & \textbf{-2.31} & -1.71  & \textbf{-2.27}  &  -1.08 & \textbf{-1.96} &  -1.14 & \textbf{14.5} & 13.4 \\ 
    \hfil{Average}  &  &  &   \textbf{-2.29}  & -1.98 &  \textbf{ -1.90} & -1.27 &  \textbf{-1.70} & -1.21 & \textbf{10.2} & 8.4 \\ 
     \hline   
     \hfil{\textit{Parkrun}}  &  \hfil{\multirow{3}[20]{*}{720p}} & 481 & \textbf{-1.49}  & -0.73  & \textbf{-0.86}  & -0.44 &  \textbf{-0.03} & 0.77 & \textbf{9.3} & 8.4 \\ 
     \hfil{\textit{Kristen and Sara}}  &   & 593 &  \textbf{2.57} & 3.74  & \textbf{0.72} & 1.20 & \textbf{0.73}  & 1.19 & \textbf{6.1} & 3.3 \\ 
     \hfil{\textit{Runners}}  &   & 289 &  \textbf{2.09} & 2.87  & \textbf{0.9} & 1.36 & \textbf{1.75}  & 1.99 & \textbf{10.9} & 9.8 \\ 
     \hfil{\textit{Into tree}}  &   & 497 &  \textbf{-1.89} & -0.73  & \textbf{-0.84} & 0.26 & \textbf{-3.03} & -0.14 & \textbf{8.4} & 6.7 \\ 
     \hfil{\textit{Old town cross}}  &   & 497 & 0.82 & \textbf{-3.59}  & \textbf{1.75}   & 2.15 & 0.92 &  \textbf{-0.07} & \textbf{7.5} & 5.3 \\ 
     \hfil{\textit{Sintel}}  &   & 337 & -2.22 & \textbf{-4.16}   & \textbf{ -1.38}   & -1.20 & \textbf{-1.59} & -0.79 & \textbf{10.9} & 8.5 \\ 
     \hfil{\textit{Rainroses}}  &   & 497 &  \textbf{0.08} &  0.52 & \textbf{0.07} & 0.56 & \textbf{-0.41} & 0.62 & \textbf{12.0} & 10.4 \\ 
     \hfil{\textit{Netflix Ritual dance}}  &   & 417 & \textbf{-2.64}  &  -1.60 & \textbf{-1.73} & -1.45  & \textbf{-2.12} & -1.64 & \textbf{9.6} & 8.2 \\ 
     \hfil{\textit{Race Night}}  &   & 593 &  \textbf{-7.13} & -5.91  & \textbf{-5.83} & -4.77  & \textbf{-5.74}  & -4.34 & \textbf{12.6} & 11.0\\ 
     \hfil{\textit{Beauty}}  &   & 593 & \textbf{-1.98} & 0.99  & \textbf{0.05}  & 0.46 &   \textbf{-0.19} & 0.54 & \textbf{7.1} & 4.7 \\ 
     \hfil{Average}   &   &  & \textbf{-1.18} & -0.86  & \textbf{-0.72}  & -0.19 & \textbf{-0.97}  & -0.19  & \textbf{9.4} & 7.6 \\  \hline
     \multicolumn{3}{|c|}{Overall average} &  \textbf{-1.73}  &  -1.42  & \textbf{-1.31} & -0.73 & \textbf{-1.34} & -0.70  & \textbf{9.8} & 8.0 \\ \hline
  \end{tabular}
  }
  \label{table:RD_performance}
\end{table*}

\begin{figure*} [htb]
    \centering
  \
       \includegraphics[width=0.49\linewidth]{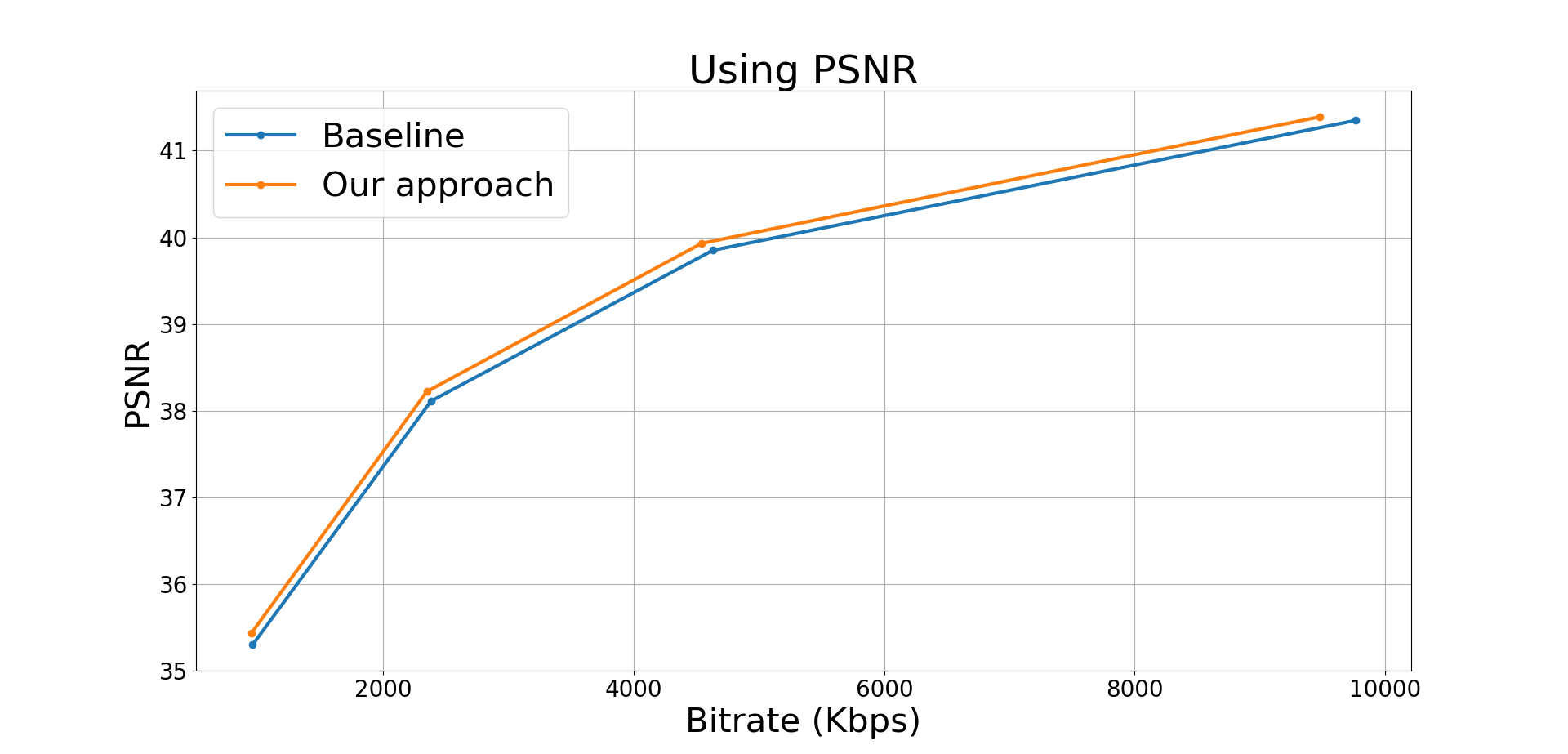}
 \hfill  
        \includegraphics[width=0.49\linewidth]{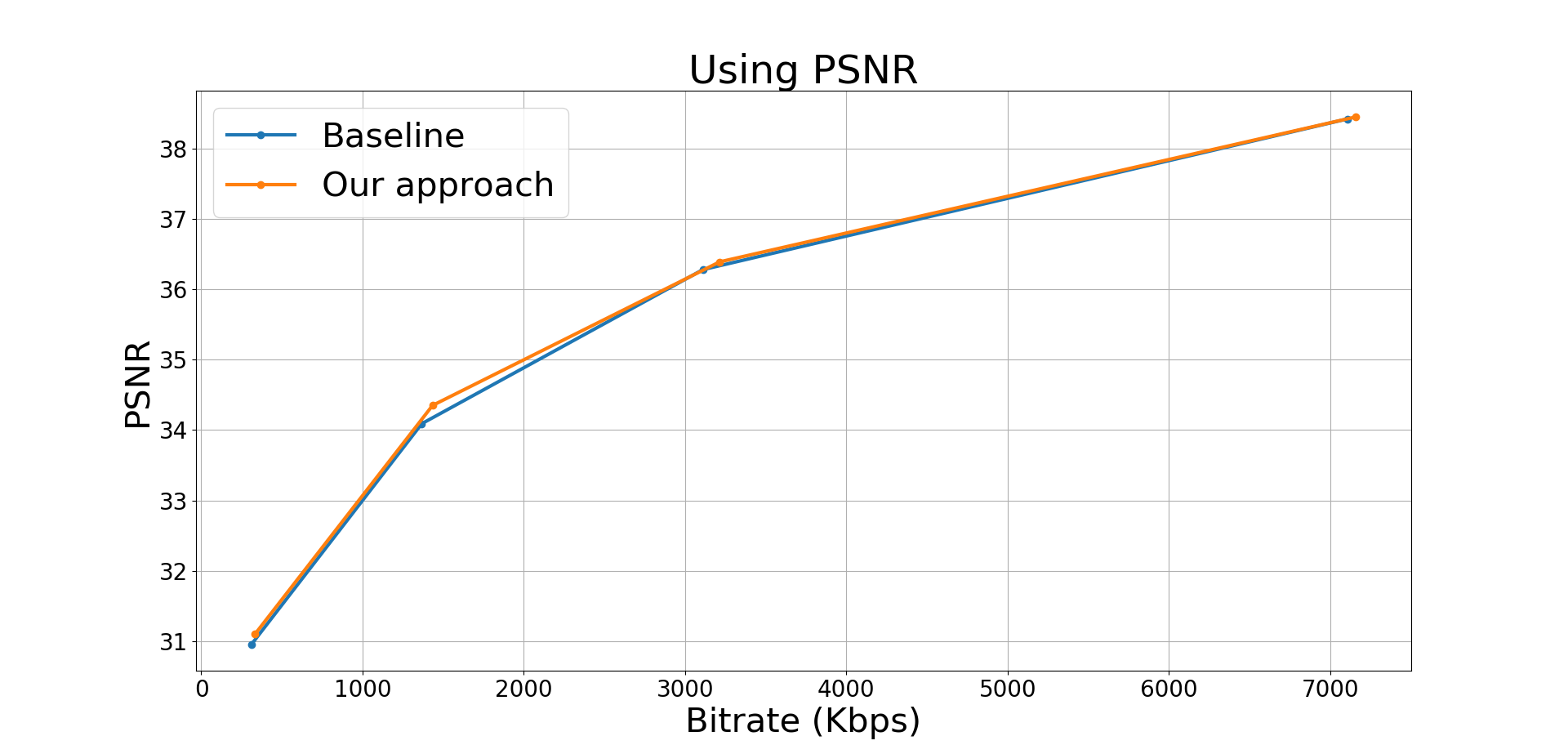}
        
   \includegraphics[width=0.49\linewidth]{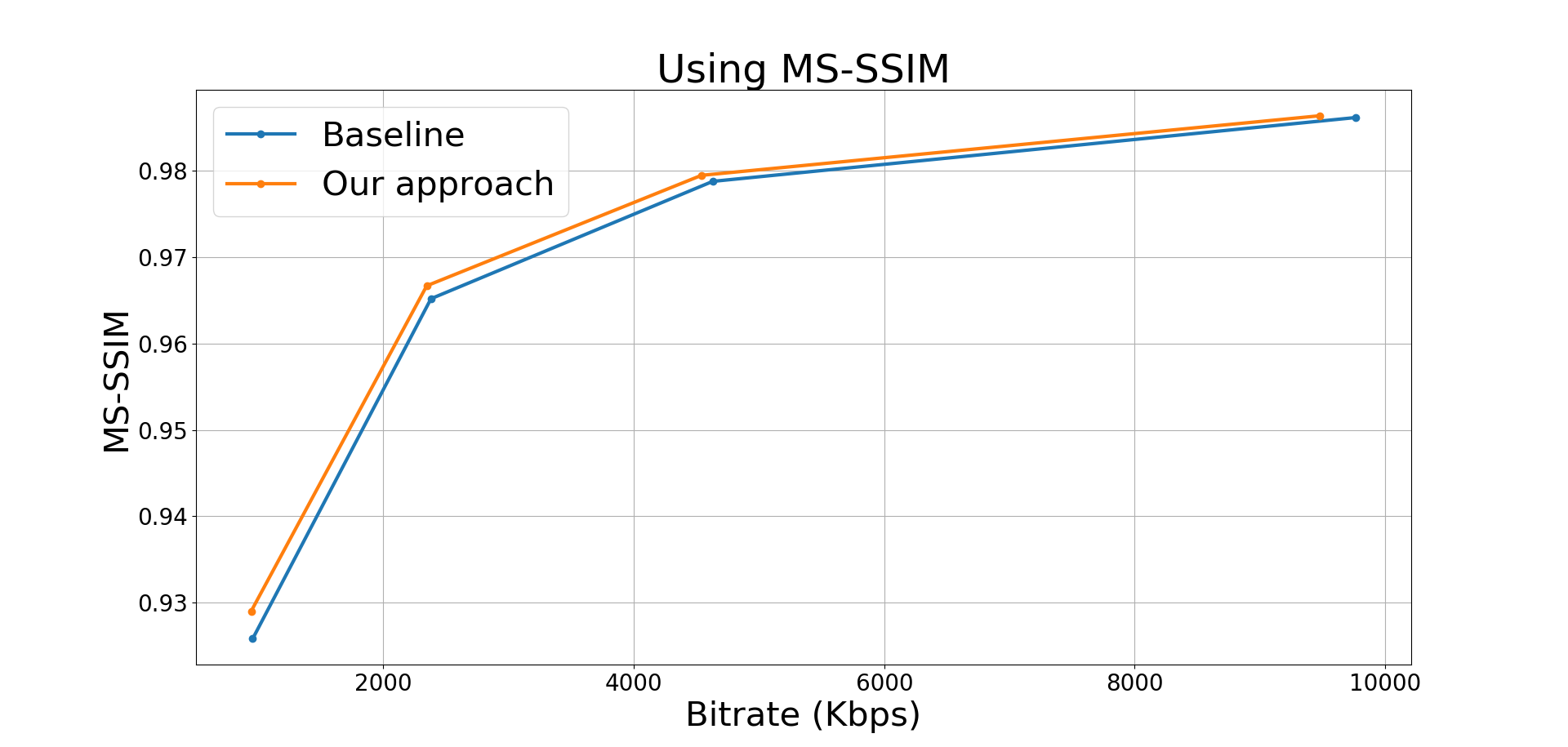}
   \hfill
  \includegraphics[width=0.49\linewidth]{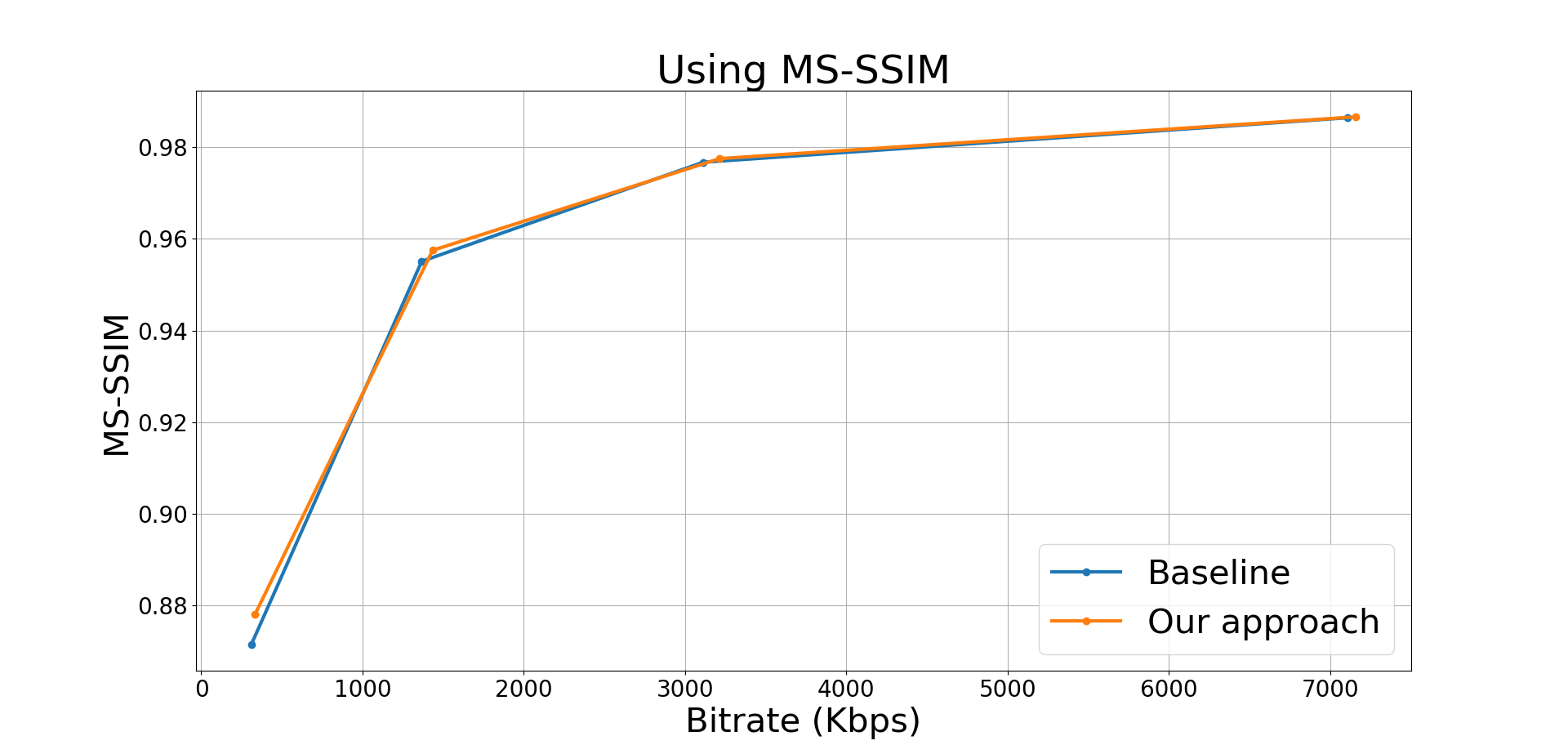}
  
   \subfloat[\textit{Touchdown pass} (1080p)\label{1b}]{%
   \includegraphics[width=0.49\linewidth]{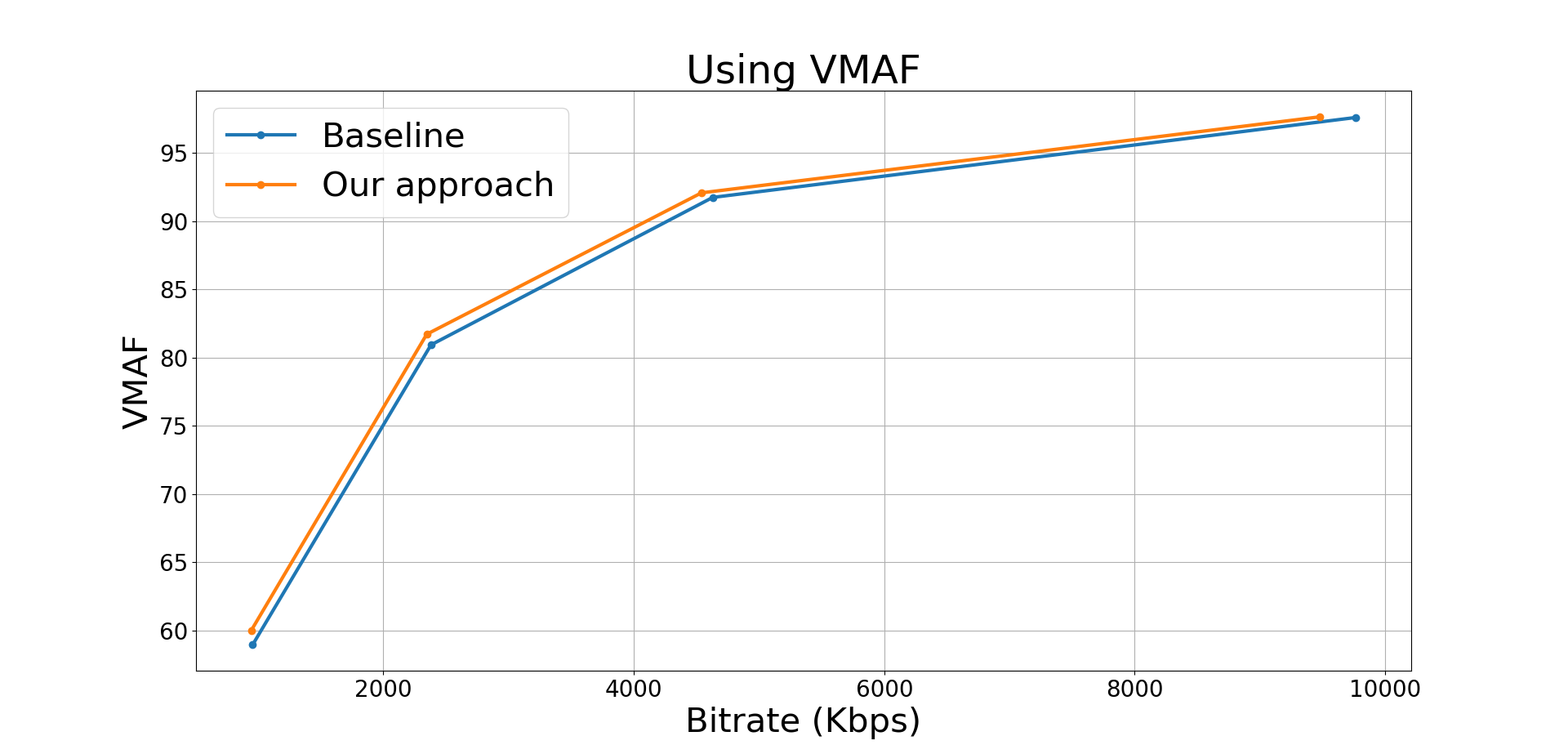}}
   \hfill
  \subfloat[\textit{Into tree} (720p)\label{1b}]{%
  \includegraphics[width=0.49\linewidth]{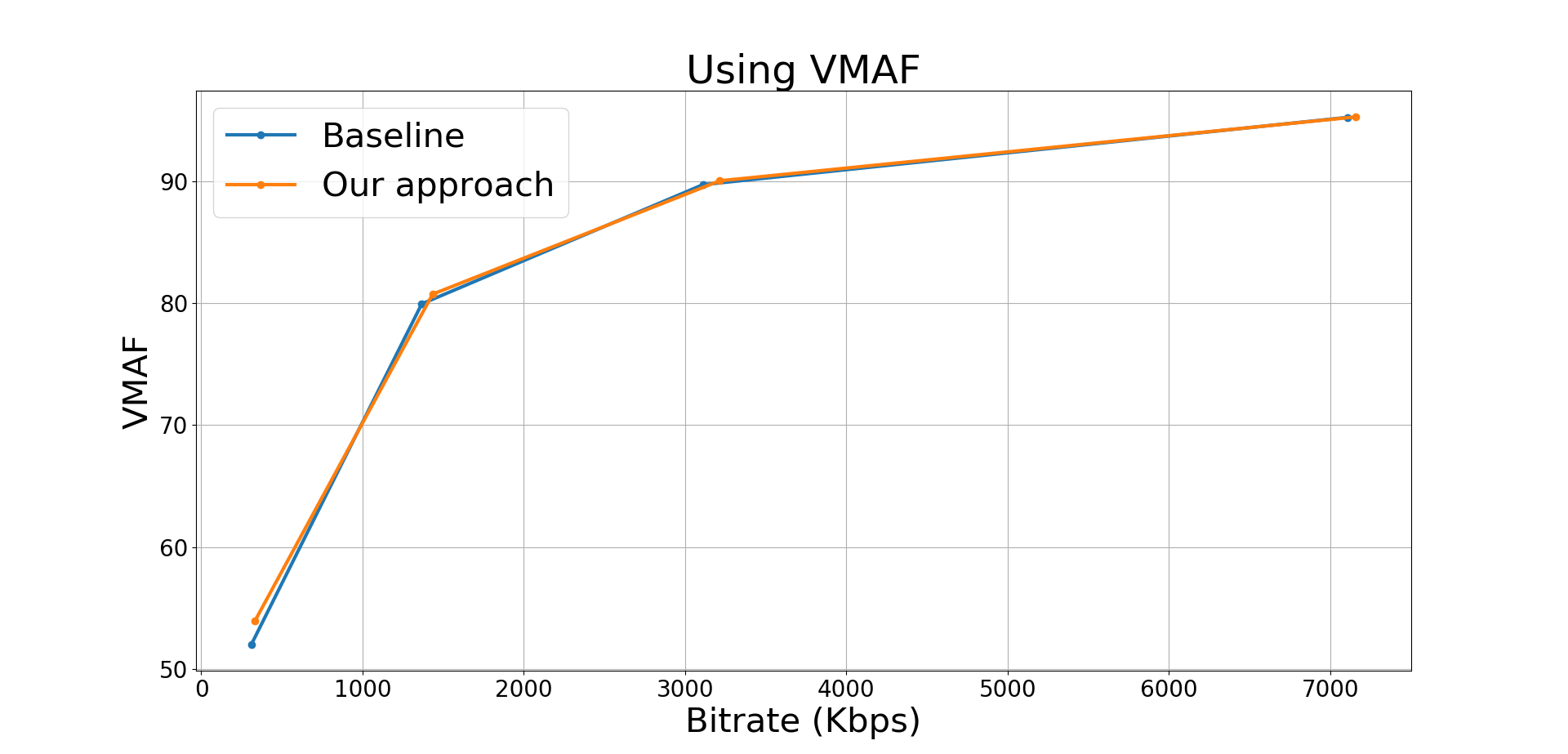}}

  \caption{RD plots of two test sequences.}
  \label{fig:RD_plots} 
\end{figure*}

\subsection{Prediction Performance}
\label{subsec:pred}
In the absence of ground truth MVs, the pixelwise error between a B-frame $Q$ and its prediction $\hat{Q}_{S \times S}^c$ acts as an approximate measure of the precision of block MVs. We computed this prediction error as the mean absolute difference (MAD) value between $Q$ and $\hat{Q}_{S \times S}^c$ for all four block sizes, on the validation partition of each of the sets $\mathcal{S}_1, \cdots, \mathcal{S}_4$. We also compared our prediction performance to the HME implementation of the SVT-AV1 encoder \cite{svt} as well as the ARPS \cite{arps} method as presented in Table \ref{table:prediction}\footnote{We used the python implementations of ARPS available in the scikit-video library at \url{http://www.scikit-video.org/stable/motion.html}, and translated to optimized machine code using Numba (\url{https://numba.pydata.org/}), although actual encoders may use more optimized implementations.}. While HME acts as the baseline for our work, the ARPS method was selected for comparison since it provides the best prediction error and computational complexity tradeoff among the widely used search pattern based block matching algorithms \cite{blockmatchstudy}. 

The average prediction errors reported in Table \ref{table:prediction} show that HME has the largest prediction errors at all block sizes and temporal layers, while ARPS achieves the smallest values. The prediction errors of the CBT-Net model are much smaller than HME and are mostly comparable to that of ARPS. 

The MVs obtained using the CBT-Net model are depicted in Fig. \ref{fig:visualization} for two sequences from the validation set, and compared against the corresponding MVs obtained using exhaustive search (ES), ARPS and HME based block matching procedures. We included the MVs given by ES in this comparison, as it produces the lowest pixelwise prediction errors among all block matching based motion estimation algorithms, although it is too computationally intensive to be practicable for evaluation on the entire validation set as given in Table \ref{table:prediction}. The MVs in Fig. \ref{fig:visualization} were color coded such that the hue and saturation represent the orientation and magnitude of the vectors, respectively \cite{floweval}. From the first three rows of Figs. \ref{fig:vis_a} and \ref{fig:vis_b}, it is evident that for all the three block matching based approaches that were considered, the estimated motion fields become progressively noisier as the block size is reduced, thereby revealing that the aperture problem is more pronounced for smaller blocks, as discussed earlier in Section \ref{subsec:arch}. In fact, in Figs. \ref{fig:vis_a} and \ref{fig:vis_b}, the ES algorithm that gives the lowest prediction errors also produces the most irregular motion fields at the two smallest block sizes as compared to the other three methods. By using the predicted motion information from larger blocks to guide estimation of the motion of smaller blocks, the CBT-Net model was able to avoid the aperture problem, yielding more coherent motion fields than the three other methods as shown in the bottom rows of Figs. \ref{fig:vis_a} and \ref{fig:vis_b}. The visual comparison in Fig. \ref{fig:visualization} shows that minimizing the pixelwise prediction error, as done by the block matching approaches such as ES, ARPS and HME was often unable to track the true motion of the objects for smaller block sizes, while our proposed multi-stage model accomplished consistent motion estimation even when block sizes were small. Since noisy motion fields are more expensive to encode in terms of the number of bits required to represent them, lower absolute pixelwise prediction errors do not necessarily ensure a better RD performance as we further demonstrate in Section \ref{subsec:rd}.

\begin{figure*}[htb]
\centering
\includegraphics[width=18cm]{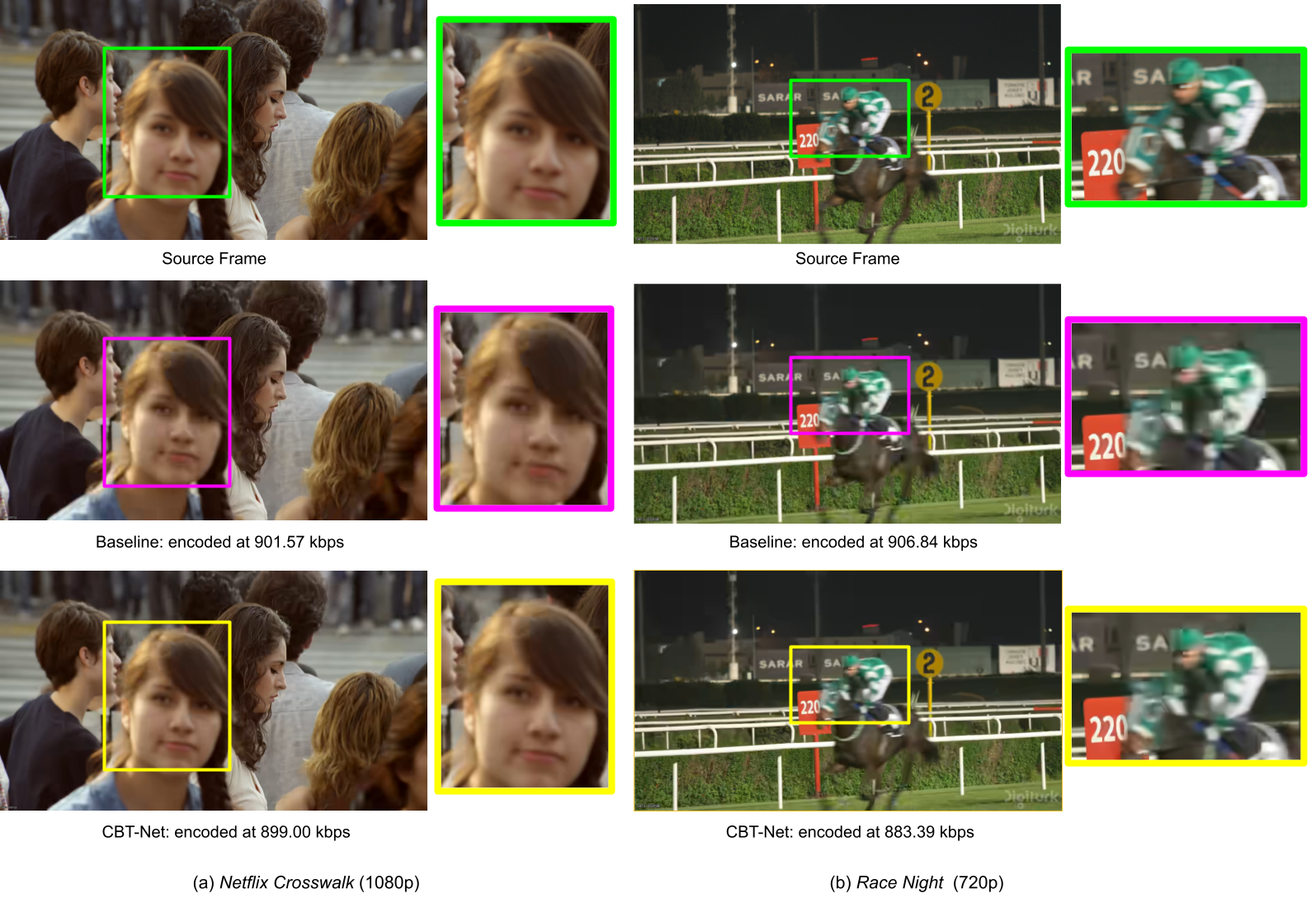}
\caption{\small{{Comparison of visual quality of the encoded frames obtained using the MVs computed by the SVT-AV1 baseline and the CBT-Net model.  }}}
\label{fig:comparison}
\end{figure*}

\subsection{RD Performance}
\label{subsec:rd}
We evaluated the RD performance of our block MV prediction approach on two resolutions (1080p and 720p), using ten test video sequences at each resolution. The contents of the test video sequences are distinctly different from the ones chosen for training and validation and they cover a wide range of motion types, as indicated by the distribution of their temporal information (TI) values (computed as the maximum over time of the standard deviation of frame differences) \cite{ti} shown in Fig. \ref{fig:ti_dist}. 

The SVT-AV1 encoder operating at speed level 1 and using HME followed by integer full search and quarter pel refinement for inter-prediction was used as the baseline for this experiment. We also configured the encoder to use the temporal prediction structure shown in Fig. \ref{fig:temporal_layers}. The test sequences were encoded at four QP values (30, 40, 50 and 60), and the B-frames were predicted using the MVs obtained in three different ways as follows:
\begin{enumerate}
\item using HME (baseline configuration of the encoder)
\item using CBT-Net
\item using ARPS
\end{enumerate}

The MVs obtained using ARPS were scaled and formatted in the same manner as described in the last paragraph of Section \ref{subsec:integration}. The integer MVs of the P-frames from temporal layer 0 of Fig. \ref{fig:temporal_layers} were computed using the integer motion search procedure of the encoder. As the mini-GOP size of the hierarchical prediction structure is 16, we encoded the first $(\lfloor K/16 \rfloor \times 16) +1$ frames of a sequence having $K$ frames, to avoid an incomplete mini-GOP at the end of the sequence. The number of frames encoded for each sequence is listed in the third column of Table \ref{table:RD_performance}. The qualities of the encodes thus obtained were evaluated using MS-SSIM \cite{ms-ssim}, VMAF \cite{vmaf} and PSNR, where the first two are perceptual quality metrics. The RD performance on the test set, as measured by the Bj{\o}ntegaard delta bitrates (BD-rates) \cite{bjontegaard} with respect to each  quality metric, are summarized in Table \ref{table:RD_performance}. 

Replacing the MVs estimated by the SVT-AV1 encoder with the CBT-Net model's MV predictions resulted in notable RD performance gains for all three metrics and at both resolutions tested, as shown in Table \ref{table:RD_performance}, while the RD plots of two test sequences are presented in Fig. \ref{fig:RD_plots}. The average BD-rates across all test sequences with respect to MS-SSIM VMAF and PSNR were -1.73\%, -1.31\% and -1.34\%, respectively, showing consistent gains across the different metrics. The RD performance gain achieved against the VMAF metric shows that the model's applicability to more general perceptual metrics, even though it was trained using MS-SSIM. This suggests that the CBT-Net learned to make MV predictions that contributed favorably towards optimizing the visual quality of the predicted B-frames. However, on some of the test sequences such as \textit{Flags}, \textit{Fountains}, \textit{Kristen and Sara}, \textit{Runners} etc. no RD gains were attained. As discussed in Section \ref{subsec:source}, the dataset contains both sequences with camera motion and sequences with little to no camera motion. Our method is advantageous, and performs especially well on sequences that contain camera motion, while little or no performance advantage is attained on sequences with primarily static backgrounds, that have few moving regions. 

Table \ref{table:RD_performance} also compares the RD performance of our method with ARPS, which achieves average BD-rates of \mbox{-1.42\%}, -0.73\% and -0.70\% with respect to MS-SSIM, VMAF and PSNR, respectively. Thus, although the ARPS method achieved slightly lower prediction errors than CBT-Net as shown in Table \ref{table:prediction}, the BD-rates it attains on the majority of the test sequences are higher than our method, substantiating our earlier claim that smoother motion fields are beneficial to attaining a better RD tradeoff at reasonably similar prediction errors. Fig. \ref{fig:comparison} compares the visual quality of the frames of two test sequences encoded with the baseline encoder against our MV estimation framework. Fewer compression artifacts are visible in the encoded frames obtained using our proposed framework, as shown in Fig. \ref{fig:comparison}. 
                 
\subsection{Encoding Complexity}
\label{subsec:complexity}
The proposed block motion estimation approach improves the RD performance of the baseline without incurring an increase in computational complexity. In fact, it reduces the computational complexity of the original SVT-AV1 encoder to some extent due to the elimination of the searches needed by HME for block matching. For each test sequence, we computed the difference in the encoding time of our method or ARPS with respect to the baseline as $\Delta T = \frac{T_0 - T}{T_0} \times 100$, where $T_0$ is the total encoding time of the baseline AV1 encoder for all QPs and $T$ is the corresponding value when the motion estimation module is substituted with our model or ARPS. The time $T$ includes the inference time of the CBT-Net model or ARPS on CPU.

\begin{table*}[htb]
  \caption{Average RD performance of CBT-Net model against PWC-Net \cite{pwc} and Unflow \cite{unflow}.}
  \centering
  \scalebox{0.95}{
  \begin{tabular}{|c|ccc|ccc|ccc|}
    \hline
    \multirow{2}{1.5cm}{\textbf{\hfil{Resolution}}} & \multicolumn{3}{c|}{\textbf{MS-SSIM BD-rates (\%)}}  & \multicolumn{3}{c|}{\textbf{VMAF BD-rates(\%)}} & \multicolumn{3}{c|}{\textbf{PSNR BD-rates (\%)}} \\
    \cline{2-10}
       & PWC-Net & Unflow & CBT-Net & PWC-Net & Unflow & CBT-Net & PWC-Net & Unflow & CBT-Net\\
    \hline
    1080p & 0.44 & 0.28 & -2.29  & 0.66 & 0.55 & -1.90 & -0.19 & -0.40   & -1.70 \\ 
    720p & 1.26 & 1.37 & -1.18  & 1.38 & 1.39 & -0.72 & 1.13 & 1.18   & -0.97 \\ 
\hline
  \end{tabular}
  }
  \label{table:comparison}
\end{table*}

The MV prediction time of the CBT-Net model is 0.14 seconds per frame, while the corresponding time for ARPS is 0.66 seconds per frame on the same CPU. ARPS takes longer to estimate the MVs as the underlying search process needs to be invoked eight times to estimate the MVs for four block sizes and two reference frames, while our method predicts all the required MVs simultaneously in a single forward pass.  The $\Delta T$ values of our approach as reported in Table \ref{table:RD_performance} shows that our approach is faster than the baseline encoder on all the test sequences, achieving an average speedup of 10.2\% and 9.4\% on the 1080p and 720p test sequences, respectively. The corresponding average $\Delta T$ values obtained using ARPS are 8.4\% and 7.6\% as also reported in Table \ref{table:RD_performance}. As the total encoding time at all QP values is much larger than the time it takes to estimate MVs using either method, we get comparable speedups using the MVs derived from CBT-Net and ARPS, with our approach being slightly faster. Thus, using the composite learned model, we were able to improve the RD performance with respect to both the baseline encoder and ARPS without having a detrimental impact on the encoding speed. 

\subsection{Comparison with Deep Motion Estimation Frameworks}
\label{subsec:comparison}
To the best of our knowledge there is no published work on deep learning based block motion estimation. Hence, to conduct a comparison with existing deep motion estimation schemes, we instead studied deep optical flow estimation methods. The following data-driven dense optical flow estimation models were selected for comparison:
\begin{itemize}
\item PWC-Net \cite{pwc} - a supervised framework
\item Unflow \cite{unflow} - a self-supervised framework. 
\end{itemize}

Since the PWC-Net and Unflow models were designed to estimate dense optical flow, we derived the per block MVs by downsampling the dense optical flow field corresponding to each of the four block sizes using bicubic resampling. The resulting motion fields were used as MVs to encode the videos, following the framework described in Section \ref{subsec:integration}. Table \ref{table:comparison} compares the average RD performance of PWC-Net \cite{pwc} and Unflow \cite{unflow} with that of CBT-Net. 

\begin{table}[htb]
  \caption{Comparison of PWC-Net \cite{pwc}, Unflow \cite{unflow} and CBT-Net in terms of computational complexity.}
  \centering
  \scalebox{1.0}{
  \begin{tabular}{|c|c|c|c|}
    \hline
    \textbf{Model} & \textbf{PWC-Net} & \textbf{Unflow} & \textbf{CBT-Net} \\ \hline
    \# of trainable parameters & 9.37M & 116.58M & 1.91M\\ \hline
    \# of FLOPS  & 771.14G & 1894.8G & 41.29G \\ \hline
     Inference time on GPU & 0.26s & 0.48s & 0.0025s \\ \hline
    Inference time on CPU & 5.56s & 8.46s & 0.14s \\

\hline
  \end{tabular}
  }
  \label{table:complexity}
\end{table}

By contrast with our method, Table \ref{table:comparison} shows that RD losses were incurred on average when the dense optical flow fields predicted by PWC-Net \cite{pwc} and Unflow \cite{unflow} were used to derive block MVs. This experimental result suggests that the dense optical flow fields predicted by these methods do not provide an efficacious alternative to derive block MVs. Furthermore, the estimation of dense optical flow is generally much more computationally intensive than block motion estimation, which makes it uneconomical to derive block MVs from dense optical flow fields. This is also demonstrated with the help of Table \ref{table:complexity}, that reports the number of trainable parameters, the number of floating point operations per second (FLOPS) as well as the inference times (for 1080p input frames) of the three models being compared. The data presented in Table \ref{table:complexity} reveal that the CBT-Net model has significantly fewer trainable parameters, FLOPs and lower inference times as compared to  \cite{pwc} and \cite{unflow}, which were designed for dense optical flow estimation.

\section{Conclusion}
\label{sec:conclusion}
We developed a composite model to collectively estimate block MVs for multiple block sizes using a multi-stage CNN. The resulting CBT-Net model was trained on a database of frame-triplets created from publicly available video sources to support a hierarchical bidirectional inter prediction structure commonly used in hybrid codecs. We trained the CBT-Net model instances using a MS-SSIM based loss function in order to favor the perceptual quality of motion compensated frame predictions over pixelwise prediction errors traditionally used as block matching criteria. The proposed framework was applied to AV1 encoding, where substituting the integer motion search of the SVT-AV1 encoder with the trained CBT-Net model resulted in average BD-rates of -1.73\% and \mbox{-1.31\%} with respect to MS-SSIM and VMAF, respectively, outperforming the corresponding gains obtained using the ARPS based block matching algorithm for motion estimation. Our  approach also attained faster encoding speeds as compared to the HME baseline and ARPS. The RD performance gain establishes the efficacy of our approach in improving the perceptual quality of the encoded videos, which is further substantiated by visual comparisons. As a pertinent future direction, the extension of our current framework to capture more complicated motion types that cannot be accounted for by  simple block translations might be considered. 

\appendices
\section{Computing the Effective Receptive Field of CBT-Net}

The effective receptive field (ERF) of a CNN is defined as the area of the input image that affects a single feature in the output feature map. The ERF progressively enlarges as the depth of the network increases. The ERF of a layer at depth $k$ can be computed recursively using the following formula \cite{erf}:
\begin{equation}
R_k = R_{k-1} + (f_k-1)\prod_{i=1}^{k-1}s_i
\end{equation}
where $R_k$  is the ERF of layer $k$, $f_k$ is the size of the convolutional kernel at layer $k$, and $s_i$ is the stride of the $i^{\text{th}}$ layer. $R_0 = 1$ at beginning, i.e. at the level of the input image. Applying this formula to the CBT-Net layer by layer, gives us an ERF of 255 at the final feature extraction layer as computed step by step in Table \ref{table:erf}. 
\begin{table}[htb]
  \caption{ERFs of CBT-Net's feature extraction layers.}
  \centering
  \begin{tabular}{|c|c|c|c|c|}
    \hline
    
     Layer \#  ($k$) & filter size ($f_k$) & stride ($s_k$) & $\prod_{i=1}^{k-1}s_i$ & ERF ($R_k$) \\ \hline
     1 & 7 & 2 & - & 7 \\ \hline
     2 & 5 & 2 & 2 & 15 \\ \hline
     3 & 5 & 2 & 4 & 31 \\ \hline
     4 & 3 & 1 & 8 & 47 \\ \hline
     5 & 3 & 2 & 8 & 63 \\ \hline
     6 & 3 & 1 & 16 & 95 \\ \hline
     7 & 3 & 2 & 16 & 127 \\ \hline
     8 & 3 & 1 & 32 & 191 \\ \hline
     9 & 3 & 2 & 32 & 255 \\ \hline
    
  \end{tabular}
  \label{table:erf}
\end{table}

\balance
\bibliographystyle{IEEEtran}


\vskip 0pt plus -1fil

\end{document}